\newif\ifTwoColumns
\TwoColumnstrue         

\ifTwoColumns
\documentclass[twocolumn]{IEEEtran}
\else
\documentclass[onecolumn,11pt,draftclsnofoot]{IEEEtran}
\fi
\usepackage{amsmath}
\usepackage{amsfonts}
\usepackage{amssymb}
\usepackage{graphicx}
\usepackage{cases}
\usepackage{subfigure}
\usepackage{cite}
\usepackage{booktabs}
\usepackage{rotating}
\usepackage{multirow}
\usepackage{comment}
\usepackage{url}
\usepackage{threeparttable}
\usepackage{pdfpages}

\newcommand{\n}{{\bf n}}

\newcommand{\eg}{{\em e.g., }}
\newcommand{\ie}{{\em i.e., }}
\newcommand{\aka}{{\em a.k.a., }}

\newcommand{\expp}[1]{e^{#1}}

\def\fmax{f_\textrm{max}}

\newcommand{\FigWidth}{1}

\newcommand{\beq}{\begin{equation}}
\newcommand{\eeq}{\end{equation}}

\newcommand{\fnyq}{{f_\textrm{NYQ}}}

\newcommand{\nth}{{\n_\textrm{thermal}}}
\newcommand{\ipt}{{\textrm{IP3}}}
\newcommand{\SNRreq}{{\textrm{SNDR}_\textrm{req.}}}
\newcommand{\AMPreq}{{\textrm{Pout}_\textrm{req.}}}
\newcommand{\fsys}{{F_\textrm{system}}}
\newcommand{\db}{{\textrm{ dB}}}
\newcommand{\dbm}{{\textrm{ dBm}}}

\title{
Xampling: Analog to Digital at Sub-Nyquist Rates
\thanks{
This work has been submitted to the IEEE for possible
publication. Copyright may be transferred without notice, after
which this version may no longer be accessible.}\thanks{
The authors are with the Technion---Israel
Institute of Technology, Haifa 32000, Israel. Emails:
moshiko@tx.technion.ac.il, yonina@ee.technion.ac.il, doleg@ee.technion.ac.il, elis@ee.technion.ac.il.}}
\author{Moshe Mishali,~\IEEEmembership{Student~Member,~IEEE}, Yonina~C.~Eldar,~\IEEEmembership{Senior~Member,~IEEE}, Oleg Dounaevsky and Eli Shoshan}

\date{\today}
\usepackage[normalem]{ulem}

\begin{document}

\includepdf[pages=1]{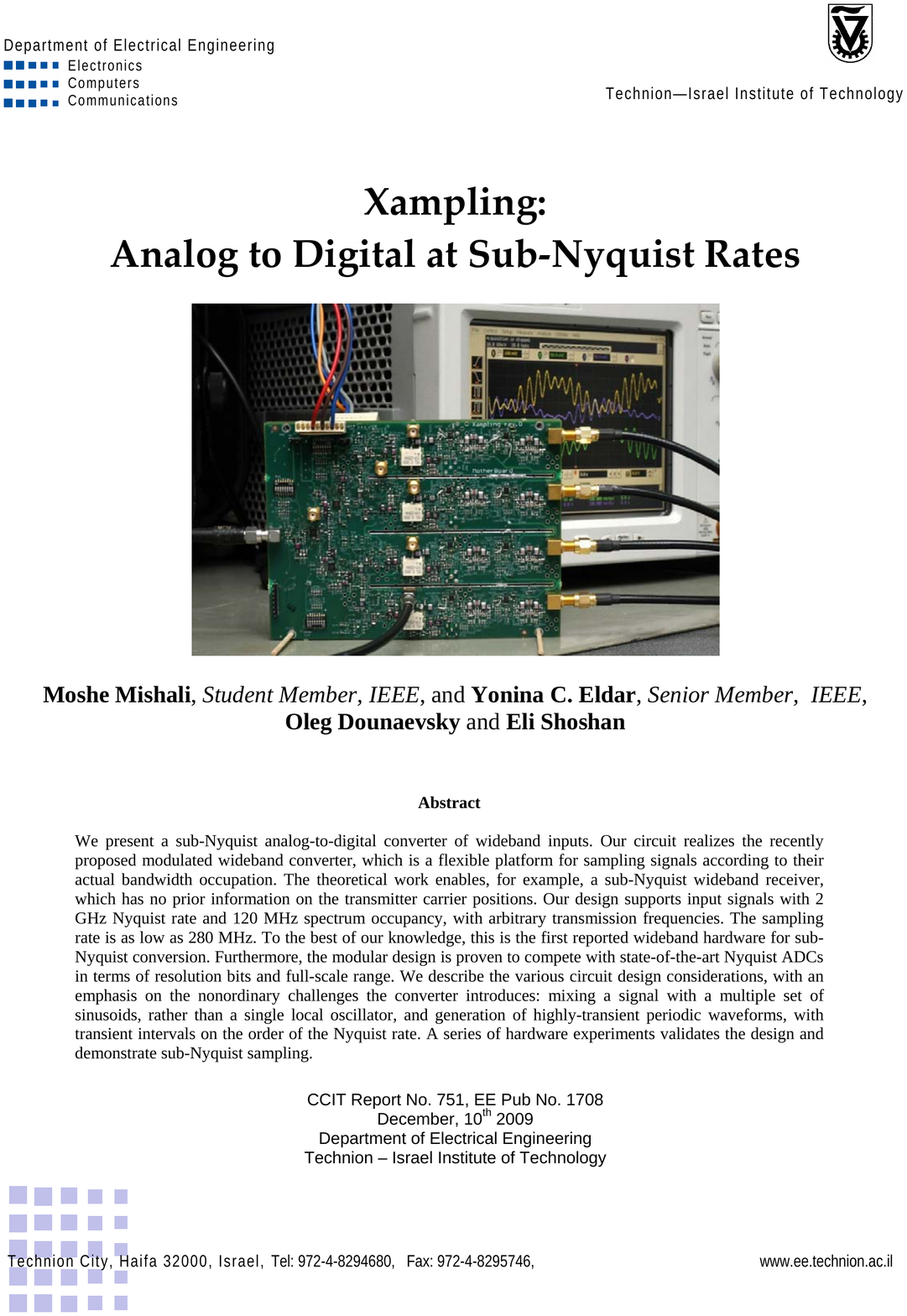}

\maketitle \IEEEpeerreviewmaketitle

\begin{abstract}
We present a sub-Nyquist analog-to-digital converter of wideband inputs. Our circuit realizes the recently proposed modulated wideband converter, which is a flexible platform for sampling signals according to their actual bandwidth occupation. The theoretical work enables, for example, a sub-Nyquist wideband receiver, which has no prior information on the transmitter carrier positions. Our design supports input signals with 2 GHz Nyquist rate and 120 MHz spectrum occupancy, with arbitrary transmission frequencies. The sampling rate is as low as 280 MHz. To the best of our knowledge, this is the first reported wideband hardware for sub-Nyquist conversion. Furthermore, the modular design is proven to compete with state-of-the-art Nyquist ADCs in terms of resolution bits and full-scale range. We describe the various circuit design considerations, with an emphasis on the nonordinary challenges the converter introduces: mixing a signal with a multiple set of sinusoids, rather than a single local oscillator, and generation of highly-transient periodic waveforms, with transient intervals on the order of the Nyquist rate. A series of hardware experiments validates the design and demonstrate sub-Nyquist sampling.
\end{abstract}

\begin{keywords}
Analog to digital conversion, circuit implementation, modulated wideband converter, sub-Nyquist sampling, Xampling.
\end{keywords}

\section{Introduction}

\PARstart{A}{nalog} to digital conversion (ADC) is the key enabling many of the advances in signal processing. An ADC device represents the continuous signal by a stream of numbers at finite resolution. Sophisticated software algorithms can then be used to manipulate the samples in order to achieve any desired processing. The Shannon-Nyquist theorem \cite{Nyquist,Shannon} lies at the heart of essentially all ADC devices. It states that signal bandlimited to $B$ Hz can be perfectly recovered from uniform samples, if the sampling rate is at least $2B$ samples/sec. Today, six decades after the formulation of this theorem by Shannon, there are various architectures for ADC design: flash, folding, pipelining, time-interleaving to name a few. However, the ultimate goal remains Nyquist sampling, that is conversion at a rate which is twice the highest frequency of the input.

\begin{figure}
\centering
\includegraphics[width=0.8\linewidth]{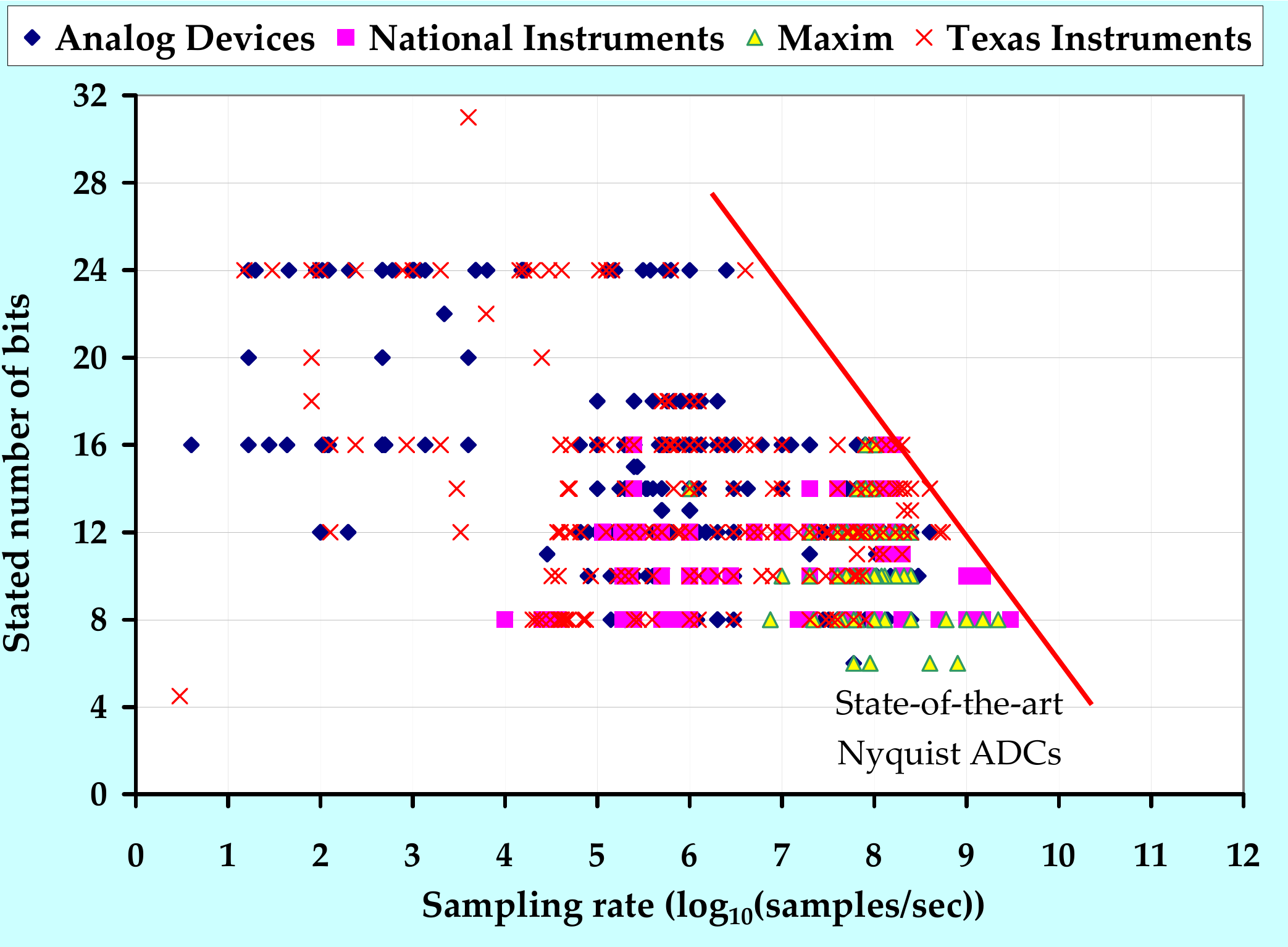}
\caption{Stated number of bits versus sampling rate. This paper demonstrates sub-Nyquist sampling of 2 GHz Nyquist-rate signals occupying only a small portion of the spectrum. The modular hardware can compete with state-of-the-art ADCs.}\label{fig:ADCstatus}
\end{figure}

A survey and analysis of ADC technology was conducted a decade ago \cite{W99} and repeated in \cite{ADCstatus05} several years ago. In Fig.~\ref{fig:ADCstatus}, we map a database of more than 1200 ADC devices from four leading manufacturers, according to online datasheets \cite{ADCratesAnalog,ADCratesNational,ADCratesMaxim,ADCratesTI}. The figure reproduces the inherent trade-off that was previously analyzed in \cite{W99,ADCstatus05}: the faster the sampling rate the smaller the number of bits that can be obtained. State-of-the-art Nyquist ADCs can achieve a sampling rate of 550 MHz at 12 bits resolution \cite{ADCratesTI}, or 3 GHz with 8 bits resolution \cite{ADCratesNational}. The expected improvement trend is an increase of 1.5 bits at all sampling rates over a period of 8 years \cite{W99}.

\ifTwoColumns \renewcommand{\FigWidth}{\linewidth} \else
\renewcommand{\FigWidth}{0.5\linewidth} \fi

The improvement trend in state-of-the-art ADCs, based on the reports \cite{W99,ADCstatus05} and Fig.~\ref{fig:ADCstatus}, is not sufficiently fast to catch up with the technology advances in related fields. For example: a prevalent scenario in communication is of a wideband receiver,
which intercepts a few narrowband transmissions, such that each
information signal is modulated on a different carrier frequency, as depicted in Fig.~\ref{fig:TypicalM}.
Nowadays, technology enables modulating information on carrier frequencies around tens of GHz \cite{VCOrates}, far above the capabilities of existing ADC devices.

The common practice in engineering to overcome the ADC bottleneck is demodulation, namely multiplying the input by the carrier frequency $f_c$ of a band of interest, so as to shift the contents of the narrowband transmission from the high frequencies to the origin. See \cite{Crols98} for common demodulation topologies. Therefore, instead of direct
sampling of a wideband input, which may be impossible with
existing technology, each band is treated individually. However, demodulation requires knowing the exact carrier frequency. This knowledge presumably requires user assistance, \eg when tuning a radio to the frequency of a station of interest. Unfortunately, such a
user-assisted solution is impractical in many applications, such
as military surveillance, radar, electronic warfare and medical
imaging. There is no time to scan for the unknown $f_c$ -- the
target may be moving and the patient cannot stand a lengthy
radiation exposure. To-date, the alternative to demodulation is to sample the entire wideband spectrum, for example by using a series of low-rate
samplers through time interleaving \cite{eldar2000frb}. This requires excessive hardware solutions with extremely high analog bandwidths, and in
addition necessitates sophisticated digital algorithms for timing synchronization and recovery \cite{huang2007blind,camarero2008mixed,ECM09,NM09,tsai2009correction,wang2006background}.

In this paper, we present a circuit-level hardware prototype of a sub-Nyquist sampling system. Our design realizes the modulated wideband converter (MWC) strategy of \cite{ME09T2P}. The MWC can sample wideband inputs at a low rate, proportional to the actual bandwidth occupation, without knowledge on the carrier positions. Specifically, our hardware can treat 2 GHz Nyquist-rate input signals with spectrum occupancy up to 120 MHz. The sampling rate can be made as low as 280 MHz, which is only 14\% of the Nyquist rate, and just slightly above the theoretical lower bound of 240 MHz average sampling rate for signals with unknown carrier positions \cite{MishaliSBR}. The dynamic range of input powers reaches 50 dB.

The MWC strategy, as briefly overviewed in Section~\ref{sec:mwc}, involves mixing the signal with highly-transient periodic waveforms, then lowpass filtering a narrowband set of frequencies followed by standard lowrate ADCs. We refer to our realization of the MWC as X-ADC, since it complies with the Xampling methodology \cite{XamplingPractice} for sub-Nyquist ADC systems. The X prefix symbolizes the rate reduction, hinting at the compression, which is conceptually carried out by analog means, concurrently with the conversion to digital. The principles of Xampling are also reviewed in Section~\ref{sec:mwc}. To the best of our knowledge, based on \cite{XamplingPractice} and the discussion in the sequel, our system is the first wideband hardware accomplishing minimal sub-Nyquist rates without knowledge of the carrier positions.

Furthermore, since the MWC strategy as well as our circuit design are modular, the approach can scale up to the Nyquist rate, providing an alternative technology for high-speed ADCs. We show that the design can reach a conversion rate of 2.075 GSamples/sec with 5.7 effective number of bits (ENOB) for a full scale range as low as 10 millivolts peak-to-peak (mVptp) inputs. To compare, MAX109 achieves 2.2 GSamples/sec with 6.68 ENOB over a 500 mVptp full scale range \cite{ADCratesMaxim}. The prime focus of the present paper is on the sub-Nyquist capabilities of our circuit design. Future works will quantify the applicability for Nyquist ADCs.

Our second contribution is the circuit solutions we incorporated in order to address the challenges that we encountered while realizing the MWC. Instead of resorting to expensive ad-hoc hardware solutions, we preferred cheap off-the-shelf devices and had to accomplish the desired functionality by wisely operating commercial devices beyond their intended specifications. The first circuit challenge is the analog preprocessing stage of the MWC, which boils down to mixing a signal with a set of multiple sinusoids. We used a standard switching-type mixer, though the ordinary use of this device, as specified in datasheets, is for multiplication with a single oscillator source, presumably for up or down conversion with a given carrier frequency. We have characterized the mixer and adjusted datasheet parameters for our usage. In addition, we added a wideband equalization preceding the mixer, and a tunable power-control circuitry. The second challenge pertains to generating highly-transient periodic waveforms that are needed for the mixing. The high-speed transients, on the order of the 2 GHz Nyquist rate, require meeting strict timing constraints. By capitalizing on the properties of specific shift-register devices we successfully managed to satisfy the strict timing requirements, with no time-delay elements, so as to avoid synchronization problems that are popular in time-interleaved ADCs \cite{ECM09,NM09}.

\ifTwoColumns \renewcommand{\FigWidth}{0.7\linewidth} \else
\renewcommand{\FigWidth}{0.5\linewidth} \fi
\begin{figure}
\centering
\includegraphics[width=\FigWidth]{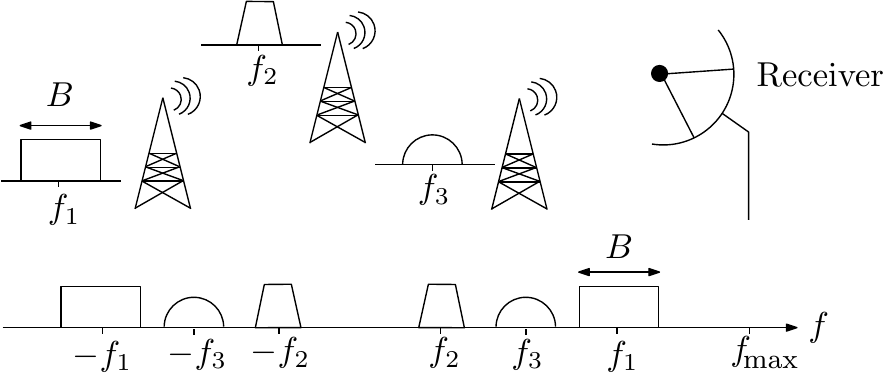}
\caption{Three RF transmissions with different carriers $f_i$.
The receiver sees a multiband signal (bottom drawing).}\label{fig:TypicalM}
\end{figure}

To verify the design, we conducted a series of lab experiments. The tests prove the periodicity of the designed mixing waveforms, which is the key for correct functionality of the MWC. Our post-manufacturing tests validate that the gains along the analog path match our preliminary theoretical computations, in which we predict the performance when taking into account the nonordinary mixing. The crowning glory of our experiments is a demonstration of sub-Nyquist sampling. We use a mixture of popular communication signals, lying anywhere in the wideband spectrum, and verify that our system aliases them to very low frequencies.

This paper is organized as follows. Section~\ref{sec:mwc} introduces the MWC system and surveys recent alternative sub-Nyquist solutions. The system specifications we chose to implement are defined and reasoned in Section~\ref{sec:spec}. The circuit design spans two sections: Section~\ref{sec:mboard} provides details on the mixing-filtering stage, whereas Section~\ref{sec:dboard} explains the periodic waveform generation. Lab experiments and reports are detailed in Section~\ref{sec:expr}. Throughout, constants are specified in either linear (\eg 1 Vptp) or logarithmic (\eg 4 dBm) formats.

\section{Theoretical background}\label{sec:mwc}

\subsection{The modulated wideband converter}

The MWC is aimed at sampling wideband sparse signals at sub-Nyquist rates assuming a multiband model. An analog signal $x(t)$ is termed multiband if its Fourier transform $X(f)$ is concentrated on $N$ frequency intervals, or bands, and the width of each band is no greater than $B$. An example is depicted in Fig.~\ref{fig:TypicalM}. The maximal possible frequency of $x(t)$, denoted by $\fmax$, dictates the Nyquist rate \cite{Nyquist,Shannon}
\begin{equation}
  \fnyq=2\fmax.
\end{equation}
The values of $N,B,\fmax$ depend on the specifications of the application at hand. In the example of Fig.~\ref{fig:TypicalM}, $N=6$ and $B$ is dictated by the widest transmission bandwidth. The multiband model does not assume knowledge of the carrier locations $f_i$, and these can lie anywhere below $\fmax$.

The MWC system has an analog front-end which preprocesses an analog multiband signal as shown in Fig.~\ref{fig:mwc}. A wideband signal $x(t)$ enters $m$ channels simultaneously. In the $i$th channel the signal is multiplied by a periodic function $p_i(t)$ with period $T_p=1/f_p$. The product $\tilde{x}_i(t)$ is filtered by an ideal lowpass filter $h(t)$ with cutoff $f_s/2$, and then sampled uniformly every $T_s=1/f_s$ seconds. In the basic configuration $m\geq 4N$ and $f_s=f_p\geq\fnyq/B$ \cite{ME09T2P}. The sampling rate is $4NB$ which can be significantly smaller than $\fnyq$. Reconstruction of the signal is obtained by appropriate digital processing, as discussed in  \cite{ME09T2P}.

\ifTwoColumns \renewcommand{\FigWidth}{0.9\linewidth} \else
\renewcommand{\FigWidth}{0.5\linewidth} \fi
\begin{figure}
\centering \mbox {
\includegraphics[width=\FigWidth]{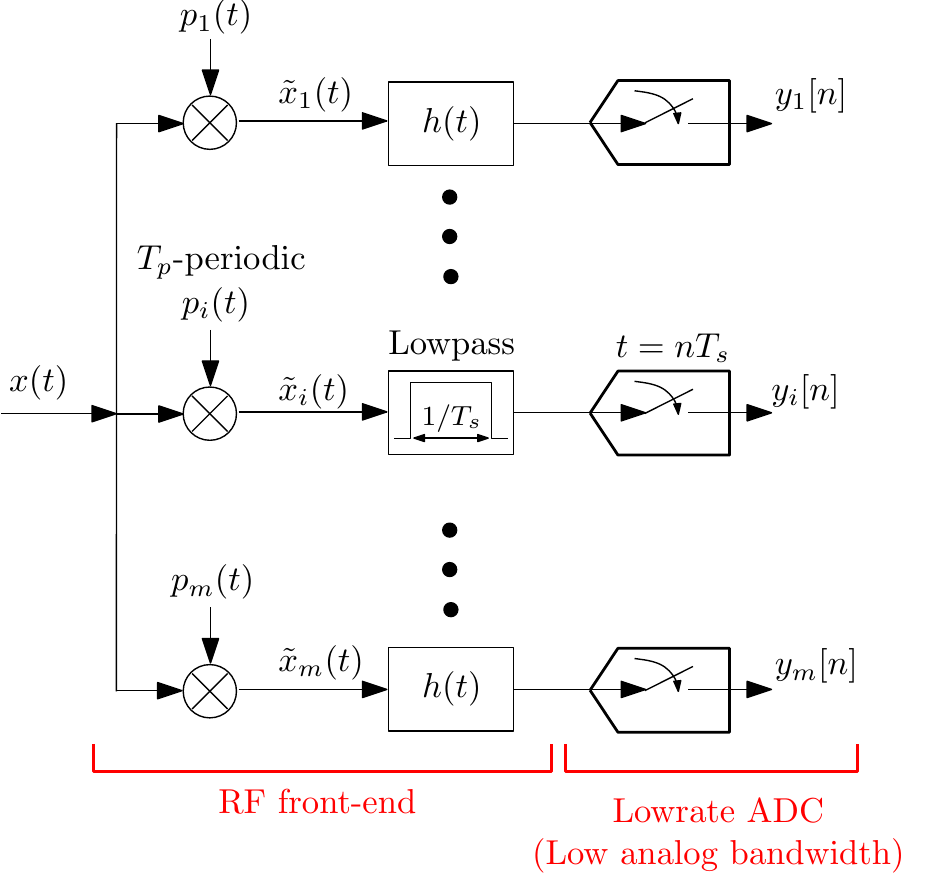}}
\caption{The modulated wideband converter: a practical sampling
stage for multiband signals.} \label{fig:mwc}
\end{figure}

The mixing operation scrambles the spectrum of $x(t)$, such that a portion of the energy of all bands appear in baseband. More specifically, since each $p_i(t)$ is periodic, it has a Fourier expansion
\begin{align}\label{eq:pitc}
p_i(t)=\sum_{l=-\infty}^\infty c_{il} \expp{j\frac{2\pi}{T_p}lt}.
\end{align}
Therefore, the mixing results in a weighted-sum of $f_p$-shifted copies of $X(f)$, such that the weights are the Fourier coefficients $c_{il}$ \cite{ME09T2P}. The lowpass filter $h(t)$ transfers only the narrow band frequencies up to $f_s/2$ from that mixture to the output sequence $y_i[n]$. The aliased output is illustrated in Fig.~\ref{fig:mixtures}. Whilst aliasing is often an undesired artifact in sampling, here it is deliberately utilized to generate mixtures at baseband. Intuitively, if $p_i(t)$ are wisely designed such that the sequences $y_i[n]$ capture different combinations of the spectrum, then the signal is determined from the samples in all branches.

Lowrate ADCs follow the RF front-end. Commercial devices can be used for that task due to the preceding lowpass. The fact that the ADCs see only a lowpass input is a major advantage with respect to time-interleaved ADCs, in which the lowrate ADCs necessitate specialized Nyquist-rate front-end since they are connected directly to the wideband input. The MWC shifts the Nyquist burden to the RF mixers, rather than imposing this complexity on the internal track-and-hold circuitry of the ADC. The low analog bandwidth ADCs and the cheap RF mixers allows realizing the MWC at a low cost. The developments in \cite{ME09T2P,XamplingPractice} provide the digital means to extract any information band of interest from the seemingly-corrupted sub-Nyquist sequences $y_i[n]$, or alternatively to recover the original input $x(t)$. Our prime goal in the present paper is to focus on the hardware realization of the RF front-end in Fig.~\ref{fig:mwc}.

\ifTwoColumns \renewcommand{\FigWidth}{0.9\linewidth} \else
\renewcommand{\FigWidth}{0.5\linewidth} \fi
\begin{figure}
\centering \mbox {
\includegraphics[width=\FigWidth]{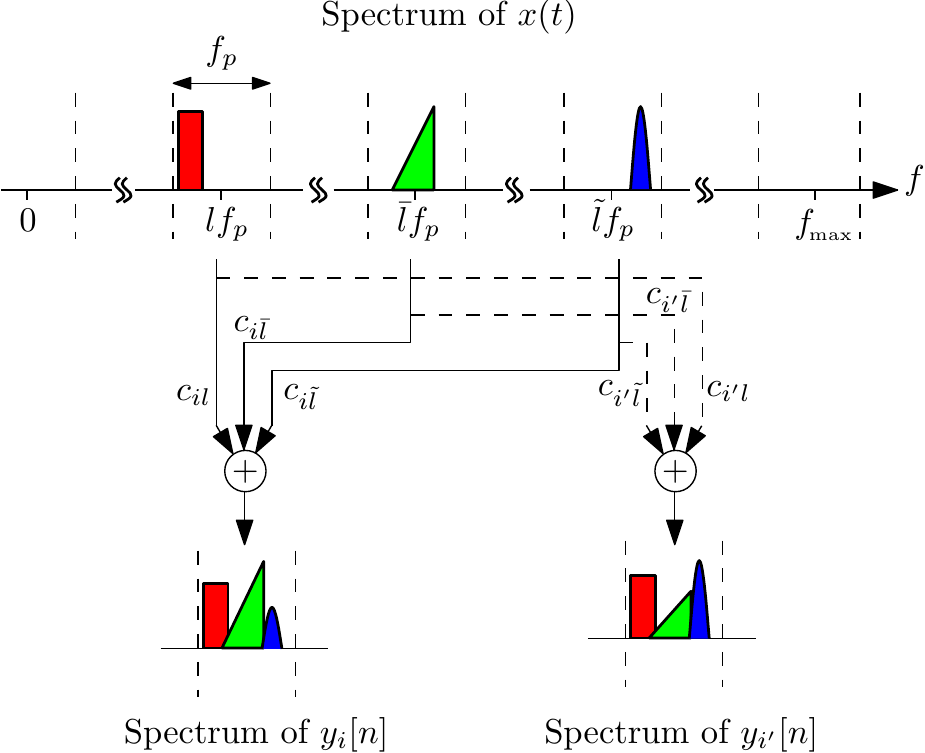}}
\caption{The spectrum slices from $x(t)$ are overlayed in the spectrum of the output sequences $y_i[n]$. In the example, channels $i$ and $i'$ realize different linear combinations of the spectrum slices centered around $lf_p,\bar{l}f_p,\tilde{l}f_p$. For simplicity, the aliasing of the negative frequencies is not drawn.} \label{fig:mixtures}
\end{figure}

The choice of the periodic functions $p_i(t)$ dictates the Nyquist rate of the input signals that the MWC can handle. In particular, if $p_i(t)$ has Fourier coefficients $c_{il}$ with non-negligible amplitudes for all $0\leq l\leq L$, then the MWC can capture signals with band locations anywhere below $Lf_p$, corresponding to a Nyquist rate of $2Lf_p$. The precise requirement on the magnitude of $c_{il}$ depends on the properties of the digital reconstruction algorithm and is beyond the current scope \cite{ME09T2P}. Nonetheless, in principle, every periodic function with high-speed transitions within the period $T_p$ can be appropriate. One possible choice for $p_i(t)$ is a sign-alternating function, with $M$ sign intervals within the period $T_p$. The choice $M\approx \fnyq/B$ was proposed in \cite{ME09T2P} and analyzed in \cite{ME09EXRIP}.

The basic MWC configuration, $f_s=f_p\approx\fnyq/B$, is sufficient for theoretical interest. In the realization of Fig.~\ref{fig:mwc}, we utilized two variants of the basic configuration, that were proposed in \cite{ME09T2P} and have practical value.
The MWC has an advanced version in which $f_s = q f_p$ for some integer $q>1$. In this setting, the sampling rate of a single channel is $q$ times higher than $f_p$. This allows to reduce the required number of channels by the same factor. Conceptually, the advanced configuration collapses the $m$ channels in Fig.~\ref{fig:mwc} at the expense of higher sampling rate per channel. Moderate values for $q$ are preferred, since this option requires additional digital computations. Another variant of interest is sharing the hardware that generates $p_i(t)$. In \cite{ME09T2P}, we used simulations to show that the reconstruction performance is hardly affected when generating several $p_i(t)$ from a single shift register (SR) of length $M$ by using different taps.

\subsection{Xampling and related works}

Xampling is a design methodology for sub-Nyquist systems which we recently introduced in \cite{XamplingPractice,XamplingTheory}. The Xampling framework aims at breaking through the Nyquist barrier by exploiting the fact that many analog signals are structured, \eg constructed from several narrow frequency bands, even though the carrier locations are unknown. Xampling requires a sub-Nyquist solution to satisfy four basic principles: capture a broad set of analog inputs, low sampling rate, efficient implementation and processing capability of any information band of interest at a low rate.

In \cite{XamplingPractice}, we compared past and recent approaches in light of the four Xampling criteria. It was shown that the Achilles heel of pointwise strategies, such as classic nonuniform sampling, is the required analog bandwidth of the ADC which remains at the Nyquist rate, regardless of the low average nonuniform rate. The analog bandwidth limitation, which stems from the speed of the ADC internal track-and-hold circuitry, holds for pointwise strategies when the carrier frequencies are known \cite{Vaidyanathan,Herley,Bresler00} and even when utilized for the case of unknown transmission locations \cite{MishaliSBR}. The literature of compressed sensing \cite{Donoho,CandesRobust} describes the random demodulation approach \cite{BN09,RDmix} which is based on discretization of the frequency axis. As discussed in \cite{XamplingPractice}, the discrete model becomes sensitive when attempting to represent continuous signals, the solution involves severe computational loads and also precludes processing at a low rate.

In the technology survey of \cite{XamplingPractice}, the MWC was found superior -- it has many advantages in a one-by-one comparison with past solutions, and the MWC is the only system to-date satisfying all four Xampling criteria. The present paper promotes the premise of the MWC from theory to practice.

We point out two related hardware publications. The technique of random demodulation underlying \cite{BN09,RDmix} was introduced in \cite{Laska07}, where the building blocks were simulated in HSPICE for synthetic inputs, matching the discretized model of \cite{BN09}. A hardware realization of the random demodulation has been presented \cite{ragheb2008prototype} for a 800 kHz Nyquist-rate input and 100 kHz sampling rate. The digital reconstruction was carried out by a 160 MHz processor. The Nyquist folding system of \cite{folding}, which was also surveyed in \cite{XamplingPractice}, was demonstrated to smear a pure sinusoid input signal into a lowrate phase-modulated output, where the input frequency can be inferred from the width of the output spectrum. The smearing effect on a narrowband transmission is more complicated, and the inference of the narrowband contents from the output was not considered in \cite{folding}. In contrast, the results we report on in Section~\ref{sec:expr} demonstrate accurate aliasing of popular communication signals from their (a-priori unknown) high-rate positions around hundreds of MHz to lowrate frequencies. The signal shape is unaltered, enabling the subsequent lowrate digital processing of \cite{ME09T2P,XamplingPractice} to extract the underlying information. To the best of our knowledge, this paper is the first to present a wideband hardware of a sub-Nyquist system.

\section{Prototype Architecture}\label{sec:spec}

The MWC is a flexible system with various parameters. In this section, we detail the specifications that we chose to realize and the main circuit challenges. Our guideline in selecting the parameters was to accomplish a system which competes with state-of-the-art ADCs, in terms of input bandwidth and resolution, while minimizing the hardware size and cost.

\subsection{Specifications}

\textbf{System scope.} Our board prototype implements the RF front-end of the MWC, from the input $x(t)$ until the analog outputs of the filters $h(t)$. This scope captures the innovative theoretical parts as well as the challenges in the actual design, which are explained in the sequel. The conversion to digital of the lowpass filtered mixtures is immediately carried out using commercial devices. In the experiments we used a four-channel scope for that purpose. 

\textbf{Input model.} We consider a wideband multiband model with $N=6$ bands, namely 3 concurrent transmissions, and maximal bandwidth $B=19$ MHz. These specifications were used in the theoretical simulations of \cite{ME09T2P} and represent a possible wideband scenario for the proof of concept. The Nyquist rate $\fnyq$ we consider is 2.075 GHz. In the prototype, this number stems from the frequency of a voltage-controlled-oscillator (VCO) we chose. Higher Nyquist rates are possible for higher VCO rates, though as $\fnyq$ increases layout design becomes a factor, beyond the proof of concept which we are interested in here.

\textbf{Periodicity.} We set the time period $T_p$ to the smallest value satisfying $1/T_p\geq B$ subject to the additional considerations which are detailed hereafter.

\textbf{Periodic waveforms.} Sign alternating waveforms at the $\fnyq$ speed are chosen, where theoretically a sign pattern of length $M\geq\lceil\fnyq/B\rceil=104$ is needed \cite{ME09T2P}. In practice, $M=108$ is used due to packaging considerations that are explained in Section~\ref{sec:dboard}. This value translates to an aliasing spacing $f_p=\fnyq/M=19.212$ MHz. To reduce the hardware size, we employed the idea of deriving all waveforms from different taps of a single SR of length-$M$. The sign values are programmable.

\textbf{Filter cutoff.} As part of the efforts to reduce hardware size, we choose the advanced MWC configuration, with $q=3$. In turn, the ideal filter cutoff would be $qf_p/2=28.5$ MHz. In practice, the filter cutoff is not sharp and exceeds to around 33 MHz. The nonideal filter response is calibrated after manufacturing \cite{Yilun09}.

\textbf{Sampling rate (per channel).} In the prototype, we sample the filter outputs with a standard four-channel scope, therefore the sampling rate $f_s$ is flexible and can approach twice the filter cutoff, $f_s=70$ MHz. When sampling the outputs with ADC devices, the sampling rate $f_s$ should be chosen in accordance to the rates of available devices. As depicted in Fig.~\ref{fig:ADCstatus}, there are many commercial ADC devices that can suit the low bandwidth of the MWC outputs. For example: 80 MSamples/sec at 16 bits resolution can be achieved by either AD9460 \cite{ADCratesAnalog}, MAX19586 \cite{ADCratesMaxim}, ADC14DS080 \cite{ADCratesNational}, or ADS5562 \cite{ADCratesTI}.

\textbf{Number of channels.} Theoretically, we need $m\geq 4N=24$ channels for the basic configuration, or $m=24/q=8$ channels for the advanced one which we use here. In the prototype, we realized only $m=4$ channels, since $y_i[n]$ are acquired by a standard four-channel scope. Compensating the factor of two is accomplished by additional digital computations \cite{MishaliSBR}. The alternative design choice of $q=6$ which would require $m=24/6=4$ channels and $f_s\approx 140$ MHz was avoided due to resolution aspects that are discussed below. Since the system is modular, a future version can duplicate the hardware and integrate a bank of 8 ADCs to eliminate additional digital computations. We note that the theory of multiband sampling necessitates a minimal sampling rate of $2NB=228$ MHz, regardless of the sampling architecture \cite{MishaliSBR}. Our realization approaches the bound with 280 MHz sampling rate, that is about 14\% of the Nyquist rate.

\textbf{Resolution.} Fig~.\ref{fig:ADCstatus} shows that 16 bit resolution is customary around $f_s\approx 70$ MHz. In comparison, for $q=6$ and $f_s\approx 140$ MHz, 12 bit resolution is more popular. Since noise always degrades the effective number of bits, we choose 16 bits.

\textbf{Circuit separation.} The design is implemented using two physical printed boards. An analog board realizes the four analog paths from $x(t)$ to the relevant filter outputs, whereas a digital counterpart provides the periodic waveforms. The separation avoids possible digital noise from interfering with the analog tasks. In addition, it allows for future designs to replace the sign alternation waveforms with any desired periodic function. Another advantage is the ability to verify each part of the design separately.

Table~\ref{table:param} summarizes the parameters for convenience.

\ifTwoColumns \newcommand{\TableSize}{} \else
\newcommand{\TableSize}{\scriptsize} \fi
\newcommand{\ra}[1]{\renewcommand{\arraystretch}{#1}}
\begin{table}\centering\caption{Prototype parameters}
\ra{1.2}{\TableSize
\begin{tabular}{@{}ll@{}}

\toprule

Parameter & Choice \\
\midrule
Signal model & $N=6$, $B=19$ MHz, $\fnyq=2$ GHz\\

Number of channels $m$ & 4 \\

Waveform type & periodic sign alternation\\

Alternation rate & 2.075 GHz \\

Sign pattern length $M$ & 108 \\

Period $f_p$ & $2.075/108 = 19.212$ MHz \\

Filter cutoff & 33 MHz\\

Sampling rate/channel $f_s$ & $70$ MHz\\
\bottomrule
\end{tabular}}
\label{table:param}
\end{table}

\subsection{Circuit challenges}

In designing an analog circuit to realize the MWC we encountered
two main show-stoppers: (I) analog mixing with spectrally-rich waveforms $p_i(t)$, and (II) constructing the periodic waveforms with the required alternation speed of 2.075 GHz. RF mixers are
tailored for multiplication with a single sinusoid, which is the standard procedure for modulating and demodulating an information band onto a given carrier $f_c$. In contrast, the MWC requires a simultaneous multiplication with many sinusoids -- those comprising $p_i(t)$. This results in attenuation of the
output and substantial nonlinear distortion not accounted for in
datasheet specifications. The next section details the design of the analog board, in which we
included a frequency equalizer
and a tunable power control on $p_i(t)$, to partially account for the nonordinary mixing. Further
research can assist in a broad investigation of mixing with multiple sinusoids, beyond the current application.

The second challenge pertains to constructing $p_i(t)$. The waveforms can be generated either by analog or digital means. Analog waveforms, such as sinusoid, square or sawtooth waveforms, are smooth within the period, and therefore do not have enough transients at high
frequencies which is necessary to ensure sufficient aliasing. On
the other hand, digital waveforms can be programmed to any desired number of alternations within the period, but require meeting timing
constraints on the order of the clock period. In our setting,
the clock interval of $1/\fnyq=480$ picosecs leads to severe timing constraints that are difficult to
satisfy with existing digital devices. Section~\ref{sec:dboard} elaborates on this challenge and the solution we selected to realize in hardware.

\section{Analog Board}\label{sec:mboard}

\ifTwoColumns \renewcommand{\FigWidth}{\linewidth} \else
\renewcommand{\FigWidth}{0.5\linewidth} \fi
\begin{figure*}
\centering \mbox {
\includegraphics[width=\FigWidth]{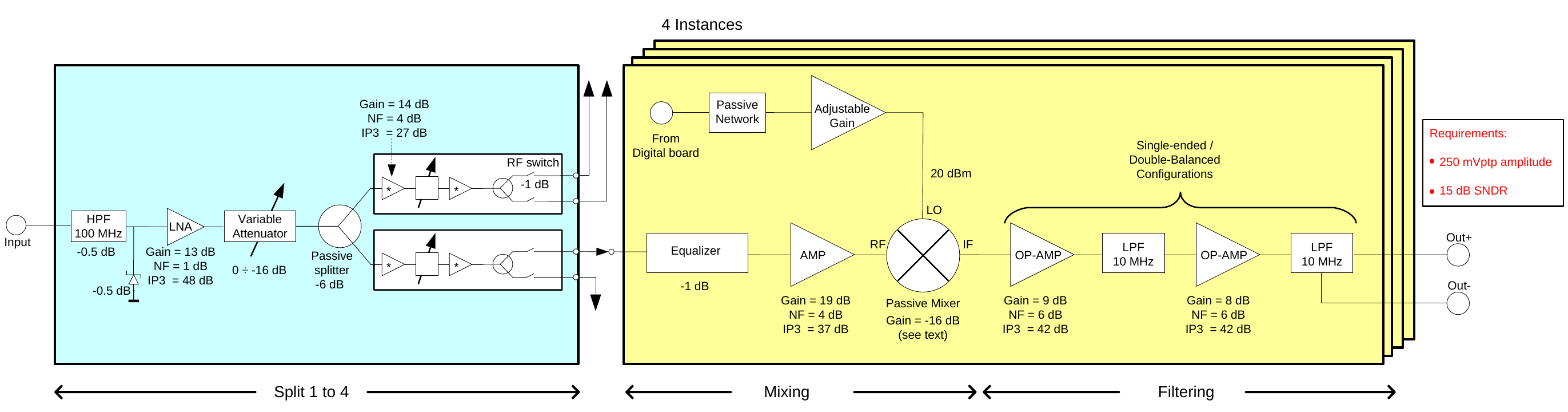}}
\caption{A block diagram of the analog board.} \label{fig:AnalogStages}
\end{figure*}

\subsection{Description}

The analog board consists of three consecutive stages: splitting the input into four channels, mixing with the sign patterns $p_i(t)$ and lowpass filtering.
Fig.~\ref{fig:AnalogStages} presents a block diagram of the analog path end-to-end.

The input signal passes through a 100 MHz highpass filter in order to reject the range of radio stations and airport transmissions, which are typically not of interest. A breakdown diode is used to protect the circuit in case of instantaneous high input power. As common in RF paths, a low-noise amplifier (LNA) leads a chain of analog amplifications. We placed several attenuators along the path, whose power loss can be digitally-controlled between 0 dB to -15.5 dB. These devices allow to widen the dynamic range of the system as explained in the next section. We point out that the actual split to four channels is carried out by splitting the signal twice. This implementation is also reasoned by dynamic range considerations. RF switches are the last components in the splitting stage. The switches allow shutting down any of the four channels, while isolating it from the active branches, so as to avoid possible RF reflections.

Four identical blocks follow. In each block, the signal is equalized and then mixed with the corresponding waveform $p_i(t)$ which is provided by the digital board. The mixing stage -- the heart of the MWC system -- is explained in Section~\ref{sec:mix}. An amplifier  mitigates possible drips of the periodic waveform into the rest of the signal path. A two-stage lowpass filtering concludes the analog path. The next subsections explain the circuit considerations underlying the structure of Fig.~\ref{fig:AnalogStages}.

\subsection{Link budget analysis}

In realizing our sub-Nyquist system, we bared in mind the representative application of a wideband receiver of Fig.~\ref{fig:TypicalM}, which intercepts several communication transmissions, appearing somewhere in a wideband spectrum. Communication signals often suffer high attenuations while in air, so that when an RF antenna intercepts the signal, the power level is typically substantially lower than the amplitude levels needed for sampling with either an ADC device or scope.  Commercial ADCs at rates around $f_s=70$ MHz can work with amplitude levels as low as tens to hundreds of millivolts \cite{ADCratesAnalog,ADCratesNational,ADCratesMaxim,ADCratesTI}. We decided to target an output amplitude of at least 250 millivolts peak-to-peak (mVptp). Equivalently, we require an output power
\begin{equation}\label{eq:ampreq}
\AMPreq\geq -8.06 \dbm.
\end{equation}
Throughout, units are exchanged according to
\begin{equation}
  \textrm{Vrms}=\frac{\textrm{Vptp}}{\sqrt{8}},\quad\textrm{Power}=10\log_{10}\left(\frac{\textrm{Vrms}^2}{Z}\right),
\end{equation}
assuming the standard $Z=50$ ohm network impedance, and that 1 dBm corresponds to 1 milliwatt. In addition, it is important to maintain a sufficiently high signal to noise ratio (SNR) so that the consecutive digital algorithms can function properly. Reconstruction in the presence of noise was demonstrated in \cite{ME09T2P}, where it was shown that an SNR of about 15 dB allows correct recovery of the spectrum support, an essential operation preceding any further processing. Since mixing is a nonlinear procedure, the output usually contains undesired spurious distorted images of the signal, which are also considered as noise. Therefore,
\begin{equation}\label{eq:snrreq}
\SNRreq \geq 15 \db
\end{equation}
stands for the required signal to noise-and-distortion ratio at the filter outputs.

Link budget calculation is a common RF analysis in order to predict the amplitude level and SNDR of an RF chain, such as Fig.~\ref{fig:AnalogStages}. The analysis takes into account the gain, noise figure (NF) and third intersection-point (IP3) of each of the devices along the path. In what follows, we analyze the chain of devices in our realization with emphasis on design considerations due to the wideband nature of the front-end. The calculations also take into account the nonordinary mixing. The experiments in Section~\ref{sec:expr} validate that the manufactured board indeed satisfies the requirements (\ref{eq:ampreq}) and (\ref{eq:snrreq}).

The amplitude requirement (\ref{eq:ampreq}) boils down to minimal input power
\begin{equation}\label{eq:pinamp}
    P_\textrm{IN}\geq \AMPreq - \sum G_i = -8.06-47 = -55 \dbm,
\end{equation}
where the gains $G_i$ are reported in Fig.~\ref{fig:AnalogStages}, and the equality above is achieved when the attenuators are set to 0 dB.

To conduct an SNDR analysis we consider two extreme scenarios. One of $x(t)$ with a very low power, in which case the SNDR is dictated mainly by the additional noise that the system generates. Technically, the noise figures and the gains of the devices along the path dictate this contribution. The other scenario assumes several narrowband transmissions, whose total power is high and is equally distributed. In this setting, spurious images of the true signal appear due to the nonlinearities of the devices. Since the input power is high, these images become dominant and distort the ability of the system to distinguish between the true input and the undesired products. In a wideband system, the effect of nonlinearities is prominent since besides the undesired products that appear close to the original frequencies, there are many other spurious images which fall well within the wideband range of the front-end.

Consider the case of $x(t)$ with low power. The thermal noise at the input of the system is
\begin{equation}
    \nth = KTW = -114 + 10\log_{10}W = -80.82 \dbm,
\end{equation}
where $KT=-114 \dbm$ is the thermal noise power per mega-Hz, and $W=2075$ MHz is the intended frontend bandwidth. It follows from Frii's formula \cite{friis}
\begin{equation}
  \fsys = \frac{\textrm{SNR}_\textrm{in}}{\textrm{SNR}_\textrm{out}}=
  F_1+\frac{F_2-1}{G_1}+\frac{F_3-1}{G_1G_2}+\cdots,
\end{equation}
that the equivalent noise figure of the system is $\fsys=3.13 \db$. Consequently, the signal power at the input must satisfy
\begin{equation}\label{eq:pinminsndr}
  P_\textrm{IN,min}\geq \SNRreq + \nth + \fsys = -62.5 \dbm,
\end{equation}
which is implied by (\ref{eq:pinamp}).

To analyze the other extreme of high input power, the attenuators are set to their maximal level of $-15.5 \db$ in order to deliberately reduce the aggregated gain, and thus decrease the nonlinear effects.
The contribution of the thermal noise for high input powers can be neglected. To derive the distortion level in this setting, we begin with the output signal power
\begin{equation}
  P_\textrm{out} = N_\textrm{max}G_\textrm{total}P_\textrm{IN,max},
\end{equation}
where $N_\textrm{max}=10$ is taken for the maximal number of concurrent narrowband transmissions that the system is designed for. Note that $N_\textrm{max}$ is greater than the 3 transmissions taken in our signal model, since the system is modular and we would like the design to support future applications with signal sets beyond the specifications of Table~\ref{table:param}. The total gain $G_\textrm{total}=\sum G_i=15 \db$ is calculated from Fig.~\ref{fig:AnalogStages}, and $P_\textrm{IN,max}$ is the unknown to be computed. We calculate the equivalent intercept point of the third order (IP3) of the entire system, using the formula \cite{IP3sum}
\begin{equation}
  (\ipt_\textrm{system})^{-1} = \sum_{i=1}^I \left(\ipt_{i}\prod_{j=i+1}^I G_j\right)^{-1}=(30.98 \dbm)^{-1},
\end{equation}
where $I=15$ is the number of RF devices in the path. Then, the SNDR at the output is given by \cite{ACPR}
\begin{equation}
  \textrm{SNDR} = 2\left(\ipt_\textrm{system}-P_\textrm{out}\right) - 7.3\db.
\end{equation}
We subtract 7.3 dB in order to account for the fact that IP3 measures the mutual distortion in the presence of two tones only. Compensating the equation for a larger number of input transmissions, $N_\textrm{max}=10$, requires this term \cite{ACPR}. Taking into account the number of concurrent transmissions is extremely important in a wideband system whose frequency bandwidth covers more than an octave. In narrowband systems, various intermodulations between any two input tones fall beyond the input bandwidth. The distortions are rejected by the system filters and thus the number of transmissions hardly affect the SNDR. In contrast, intermodulations in a wideband system fall within the input bandwidth, modulates with the other tones, and thus raises the total distortion level. We conclude that the signal input power must satisfy
\begin{align}\label{eq:pinmaxsndr}
   P_\textrm{IN,max}  \leq & \ipt_\textrm{system}-\frac{1}{2}(\SNRreq+7.3)\\&-10\log_{10}N_\textrm{max}-G_\textrm{total}=-5.17 \dbm.\nonumber
\end{align}

\begin{figure}
\centering \mbox {
\subfigure[]{\includegraphics[height=32mm]{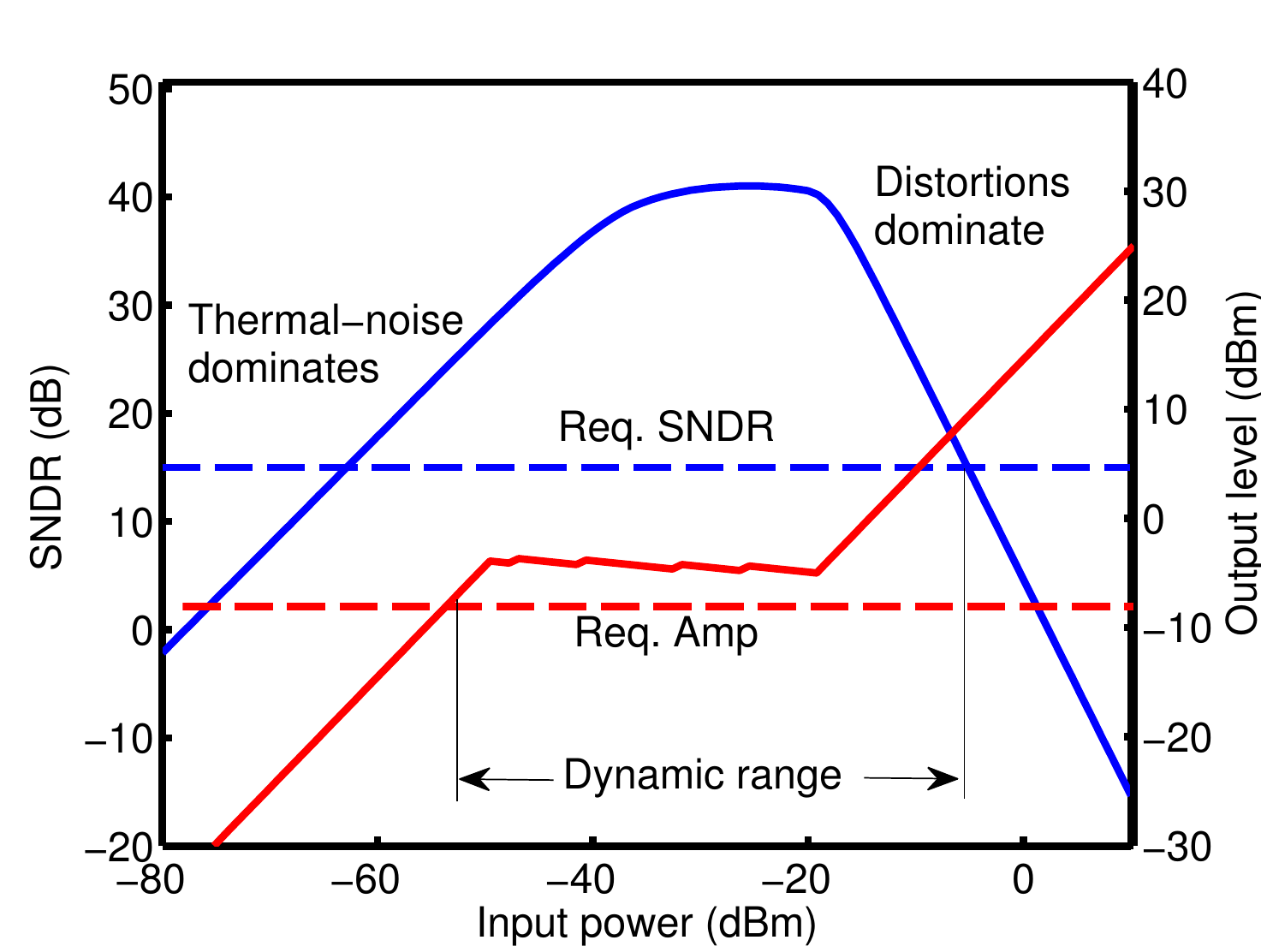}}
\subfigure[]{\includegraphics[height=32mm]{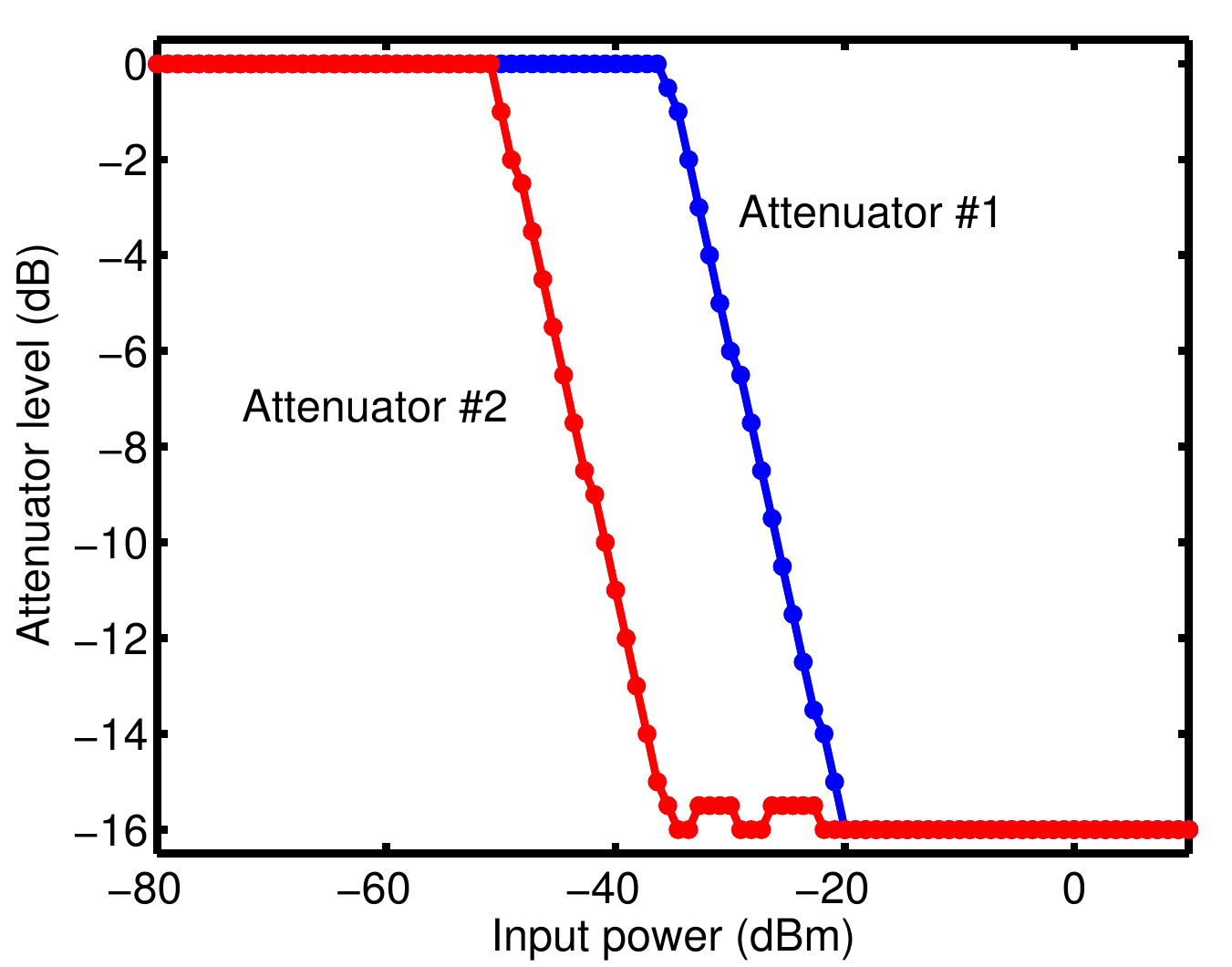}}}
\caption{The output power level and the SNDR are plotted in (a) for varying input power (solid lines), where the former is monotone in the input power. The dynamic range of the system is achieved when both requirements are satisfied. The optimal attenuator levels (in the SNDR sense) are drawn in (b).} \label{fig:sinad}
\end{figure}

Our system can therefore treat input signals at a dynamic range of 49 dB, as follows from (\ref{eq:pinamp}), (\ref{eq:pinminsndr}) and (\ref{eq:pinmaxsndr}). We note that the calculations of the distortion level for high-power signals assume that all narrowband transmissions have equal powers. Refining the theoretical calculations for different power levels is possible. However, since the link budget analysis has spares we did not pursue this direction; Section~\ref{sec:expr} reports on an actual 50 dB dynamic range of the manufactured system. Fig.~\ref{fig:sinad} depicts the output level and the SNDR for varying input power, when optimizing for the highest SNDR over all possible attenuator levels using (\ref{eq:pinminsndr}) and (\ref{eq:pinmaxsndr}). We point out that the increase in the attenuators level does not breach (\ref{eq:ampreq}) for all input power higher than -55 \dbm.

\ifTwoColumns \renewcommand{\FigWidth}{0.5\linewidth} \else
\renewcommand{\FigWidth}{0.5\linewidth} \fi

We also considered an alternative design in which the signal is split into four channels immediately after the first attenuator, with no additional splitting at the end. Repeating the link budget analysis shows that the dynamic range due to the SNDR requirement, namely the difference between (\ref{eq:pinminsndr}) and (\ref{eq:pinmaxsndr}), would have been narrowed by 4 dB, or equivalently a factor of 2.5 in the range of allowed peak-to-peak input amplitudes. When taking into account the amplitude requirement the dynamic range is narrowed by only 1 dB. Nonetheless, we avoided this option in order to improve the SNDR in case (\ref{eq:ampreq}) is relaxed in a future system. Another reason is that a direct splitting system would require an additional four amplifiers and two digital attenuators.

Fig.~\ref{fig:sinad} can be used to study the potential of the X-ADC architecture as an alternative to existing Nyquist ADCs when using $m$ branches such that $mf_s\geq \fnyq$. Specifically, Fig.~\ref{fig:sinad}(a) predicts a maximal SNDR level of 42 dB, which is achieved for an input power of -25 dBm. In practice, the experiments in Section~\ref{sec:exprsndr} affirm a maximal SNDR of 36.2 dB at -35 dBm input power. In ADC terminology, the ENOB of our system is \cite{demler1991high}
\begin{equation}\label{eq:enob}
  \textrm{ENOB}=(\textrm{SNDR}-1.76)/6.02=5.7 \textrm{ bits},
\end{equation}
for a full-scale range of 10 mVptp. Since the MWC design is modular and can scale up to the Nyquist rate, in principle, the system can reach a sampling rate of 2.075 GSamples/sec with the above specifications.

\subsection{Mixing with multiple sinusoids}\label{sec:mix}

\ifTwoColumns \renewcommand{\FigWidth}{0.8\linewidth} \else
\renewcommand{\FigWidth}{0.5\linewidth} \fi
\begin{figure*}
\centering \mbox {
\includegraphics[width=\FigWidth]{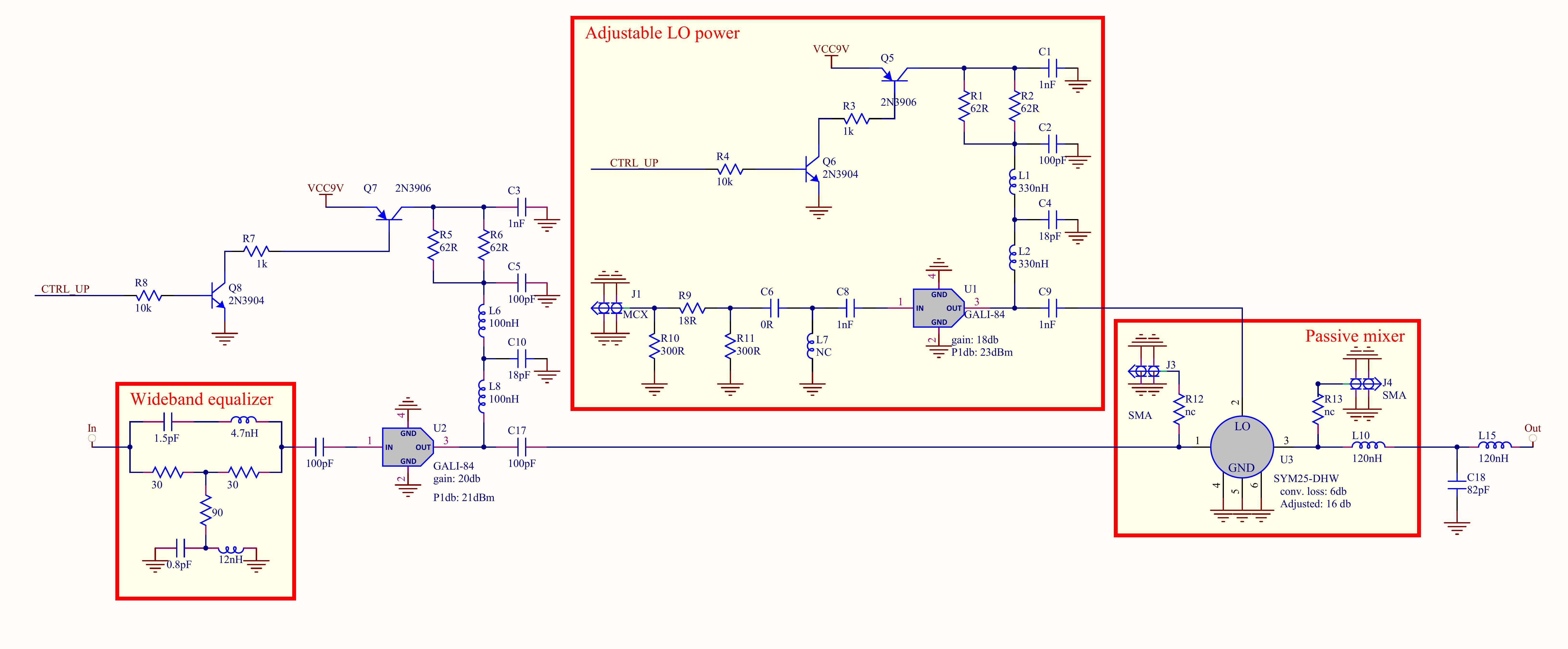}}
\caption{The schematic of the mixing stage highlights three impacts of the nonordinary mixing with multiple sinusoids: a wideband equalizer preceding the mixer, a tunable power control on LO port and a passive-type mixer. The datasheet specifications of the mixer were modified as explained in the text.} \label{fig:mixer}
\end{figure*}

The mixing stage is a major circuit challenge in realizing the MWC. A standard switching-type mixer has three ports: RF, local oscillator (LO) and intermediate frequency (IF), hinting at the common usage -- multiplication of an RF signal by a single sinusoid. In contrast, the MWC requires the nonordinary mixture $x(t)p_i(t)$, that is with the multiple sinusoids comprising $p_i(t)$. The immediate consequence of this nonordinary mixing is a choice of passive device. Active mixers, namely those with external power supply, typically have a narrow input bandwidth which is not suited for wideband sampling. The passive mixer device in our system (SYM25-DHW) is specified for RF inputs from DC to 6 GHz. The datasheets also allow LO frequency between 80 MHz and 2.5 GHz, referring to a single oscillation source. There are several other impacts on the RF path due to the nonordinary mixing of the MWC. First, in the calculations of the link budget before, we had to adjust the mixer parameters (gain, noise figure and IP3) since conventional datasheets characterize the device only for multiplication with a single LO source. These modifications impacted the hardware design as the actual gain of the system is less than expected from datasheets and additional amplifications were required. Second, we inserted an adjustable power control on the energy of $p_i(t)$ close to the LO port of the mixer. The third circuit is a wideband equalizer which we located nearby the RF port. These modifications are highlighted in Fig.~\ref{fig:mixer} and are next explained.

Link budget analysis is based on datasheet specifications (gain, noise figure and IP3). When mixing with multiple sinusoids, interpreting the device specifications become cumbersome. For example: a passive mixer requires a minimal LO power to bias the circuit; 17 \dbm~in the case of the device we chose. Here, this datum raises an interesting question: how should the LO power be set when its power is divided between the various frequencies consisting $p_i(t)$. A na\"{\i}ve approach would be to allocate a power of 17~\dbm~per harmonic, expecting a linear superposition behavior at the output. However, the mixer is a nonlinear device and the huge total power of that strategy would probably damage the device. On the other hand, allocating a total power of 17~\dbm, will result in each harmonic receiving significantly lower energy power.

In order to estimate the correct mixer parameters for the MWC setting, we performed experiments, similar to those conducted by the manufacturer, to characterize the conversion loss and the effective IP3 under a representative scenario of mixing with multiple sinusoids. We concluded that the mixer parameters should be set to conversion loss of -16 dB, IP3 of 27 dBm, and LO power of 20 dBm, instead of the datasheet specifications of -6 dB, 30 dBm, and 17 dBm, respectively. Since the experiments were conducted before manufacturing we also added an adjustable power control on the LO port of each mixer. In Section~\ref{sec:expr}, these modifications are validated on the manufactured board. We point out that RF simulation softwares, \eg SPICE or ADS, cannot accurately model the nonordinary mixing, since manufacturers provide model files only for the conventional usage.

\ifTwoColumns \renewcommand{\FigWidth}{0.7\linewidth} \else
\renewcommand{\FigWidth}{0.5\linewidth} \fi
\begin{figure}
\centering \mbox {
\includegraphics[width=\FigWidth]{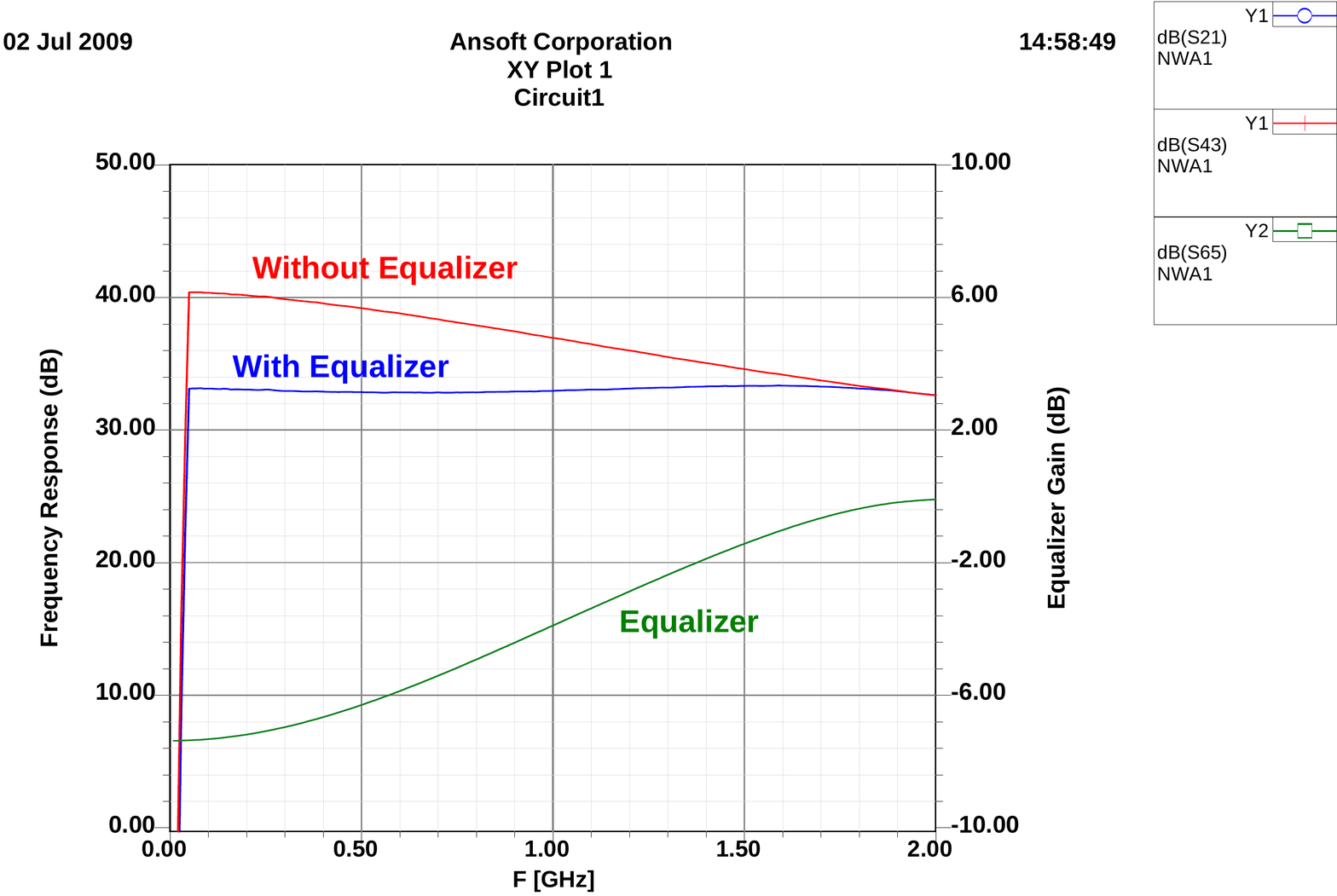}}
\caption{The wideband equalizer flattens the frequency response at the RF port of the mixer. Low
frequencies (up to 500 MHz) are attenuated by approximately 8 dB,
whereas high frequencies (above 1.5 GHz) are not altered.} \label{fig:equalizer}
\end{figure}

Moving on to the wideband equalizer. The purpose of an equalizer is to flatten the frequency response of the preceding stages. In selecting the equalizer, we had to make two uncommon choices due to the nonordinary mixing. The first regards positioning the circuit in the middle of the RF chain, just before the mixer, instead of the common knowhow of equalization towards the end of the chain, near the ADC terminals or even in the digital domain, where it can flatten the response of the entire analog path. A passive equalization, rather than an active one (\ie a circuit involving an operational amplifier), is the second choice. This selection may seem counterintuitive, since the insertion loss of a passive equalizer is high, especially since we located the equalizer in the middle of the RF chain, where the signal power is still moderate.

To explain both choices, we first recall how equalization is designed in conventional settings, that is when $x(t)$ is mixed by a single sinusoid $f_c$. RF devices attenuate the signal as it passes through, where the attenuation is frequency-dependent and tends to increase at high frequencies. The goal of the equalizer is to correct the aggregated attenuation. The design requires computing the frequency response from input to the equalizer, which as mentioned, is typically located at the end of the RF chain. Due to the frequency conversion by $f_c$, the responses of the devices preceding the mixer need to be translated by $f_c$ at the output of the mixer. The product of the responses along the path gives the aggregated attenuation, which the equalizer design fixes by adding gain or attenuation where needed, until the combined response is flattened.

In our sub-Nyquist system, the product $x(t)p_i(t)$ contains a weighed-sum of $f_p$-wide spectrum slices of $x(t)$; see Fig.~\ref{fig:mixtures}. In particular, the output contains overlayed energy from the entire signal spectrum. Consequently, there is no simple way to translate the frequency response of the devices preceding the mixer to an equivalent response at the output. An analog equalizer, if located after the mixer, cannot be designed to flatten the response of the weighed-sum in $y_i[n]$. A tricky digital equalization is possible, though at the expense of high computational loads. This explains why we placed the equalizer circuit before the mixer. Compensating for the rest of the path, namely for the nonideal response of the filter $h(t)$, is done digitally \cite{Yilun09}. The choice of passive equalizing is another consequence of the product $x(t)p_i(t)$. An active stage, which involves amplifiers, is typically limited in the frequency range that can be corrected. In our setting, the equalizer is located before the mixer, at a point that a wideband response should be fixed. In addition, an active mixer can generate nonlinear effects and we preferred to avoid adding those to the nonordinary mixture. To conclude, Fig.~\ref{fig:equalizer} displays a simulation report from the board input to the RF port of the mixer, ensuring an equalized flat response over the wideband regime until 2 GHz.

\subsection{Lowpass filtering of spectrum mixtures}

The filtering is the last processing step in the analog domain. The design focus here, which also stems from the nonordinary mixing, is on sharp transition around the cutoff and high attenuation in the stopband, namely for the frequency range beyond the cutoff. A nonflat response in the passband is not an issue, as it can be compensated digitally \cite{Yilun09}. In typical RF chains, the stopband contains mainly noise, which the filter needs to reject. Since the noise is anyway smaller in magnitude than the in-band signal level, moderate attenuation is sufficient. In contrast, the MWC is based on deliberately introducing aliasing of the signal power over the entire wideband range. Therefore, the filter is required to isolate a specific working band $[-f_s/2,+f_s/2]$, while adjacent frequencies contain non-negligible energy due to other aliasing and must be effectively eliminated.

\begin{figure}
\centering \mbox {
\includegraphics[width=0.9\linewidth]{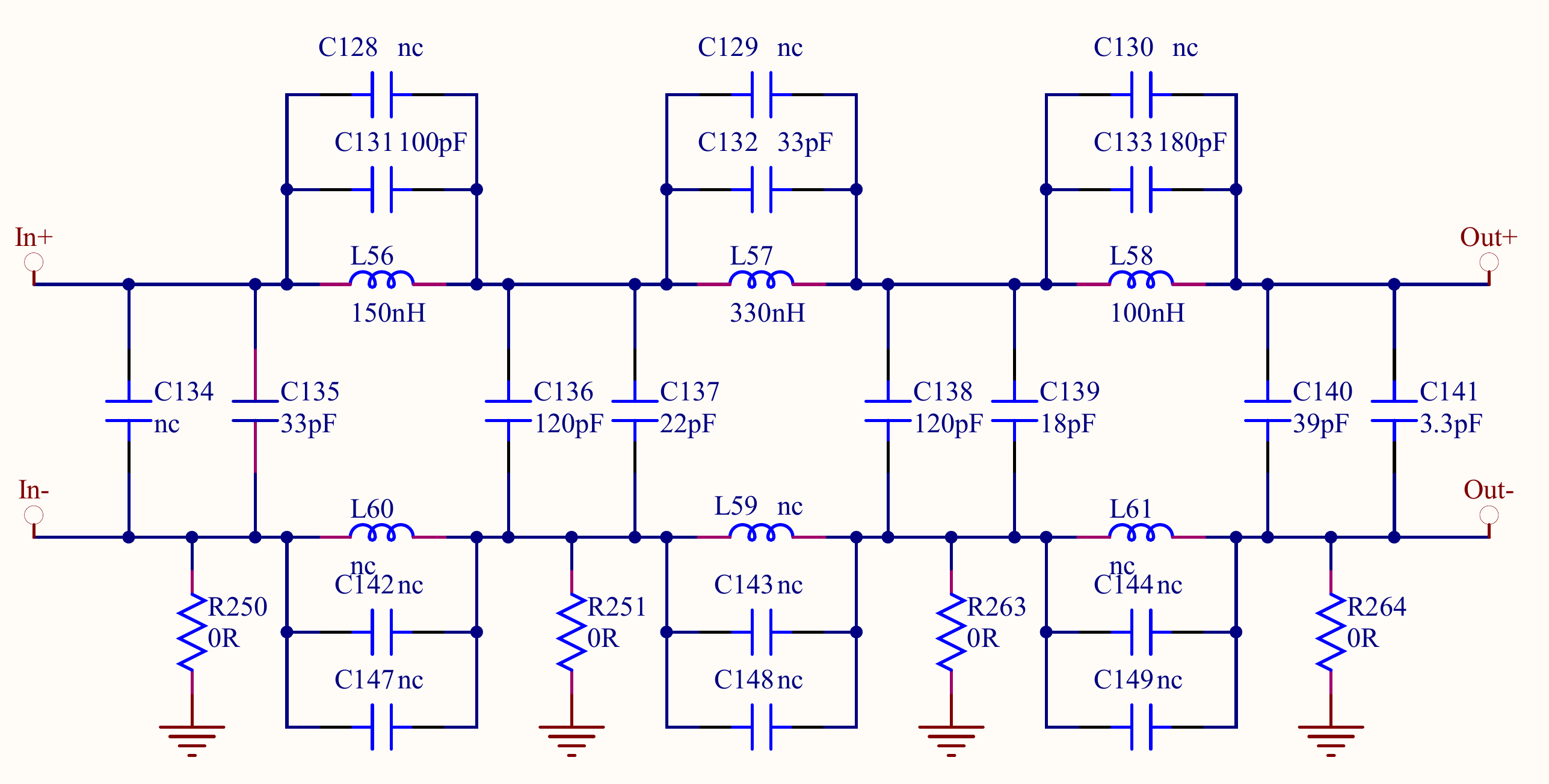}}
\caption{Lowpass filter design with 33 MHz cutoff. The component values are optimized for the single-ended output choice, in which case the parts with the `nc' designation are not assembled.} \label{fig:lowpass}
\end{figure}

To achieve the high attenuation in the stopband, we have designed an elliptic filter of order seven as appears in Fig.~\ref{fig:lowpass}. We used this filter twice with a buffering stage, as depicted in Fig.~\ref{fig:AnalogStages}, in order to double the attenuation in the stopband. Fig.~\ref{fig:LPF_Sim} presents a simulation report of the two-stage response, where we tuned the optimizer for a smooth transfer response, as measured by the curve of the $S_{21}$ parameter. We avoided a single higher-order filter due to the following reason. The $S$-parameters of a passive network satisfy $|S_{21}|^2+|S_{11}|^2=1$. The simulation report implies that the optimal $S_{21}$ is achieved at the expense of an oscillatory trend in the reflection parameter $S_{11}$. In practice, a physical network of passive elements is likely to exhibit a smooth response for $S_{11}$ as well, which in turn means that it will be difficult to obtain the optimal $S_{21}$. The oscillations in $S_{11}$ increase with the filter order. Adding to that the fact that passive elements have tolerances, which cannot be accurately modeled in simulation, led us to the design of two short-length stages.

\ifTwoColumns \renewcommand{\FigWidth}{\linewidth} \else
\renewcommand{\FigWidth}{0.5\linewidth} \fi

The filter design is flexible and allows cutoff frequencies up to 100 MHz by re-optimizing the values in Fig.~\ref{fig:lowpass}. In addition, the design supports both single-ended and double-balanced output configurations. For verification with a scope, the single-ended version is used as it matches the scope front-end. Future applications can interface our board to commercial ADCs, where balanced outputs may be preferred due to their increased immunity to noise.

\ifTwoColumns \renewcommand{\FigWidth}{0.7\linewidth} \else
\renewcommand{\FigWidth}{0.5\linewidth} \fi
\begin{figure}
\centering \mbox {
\includegraphics[width=\FigWidth]{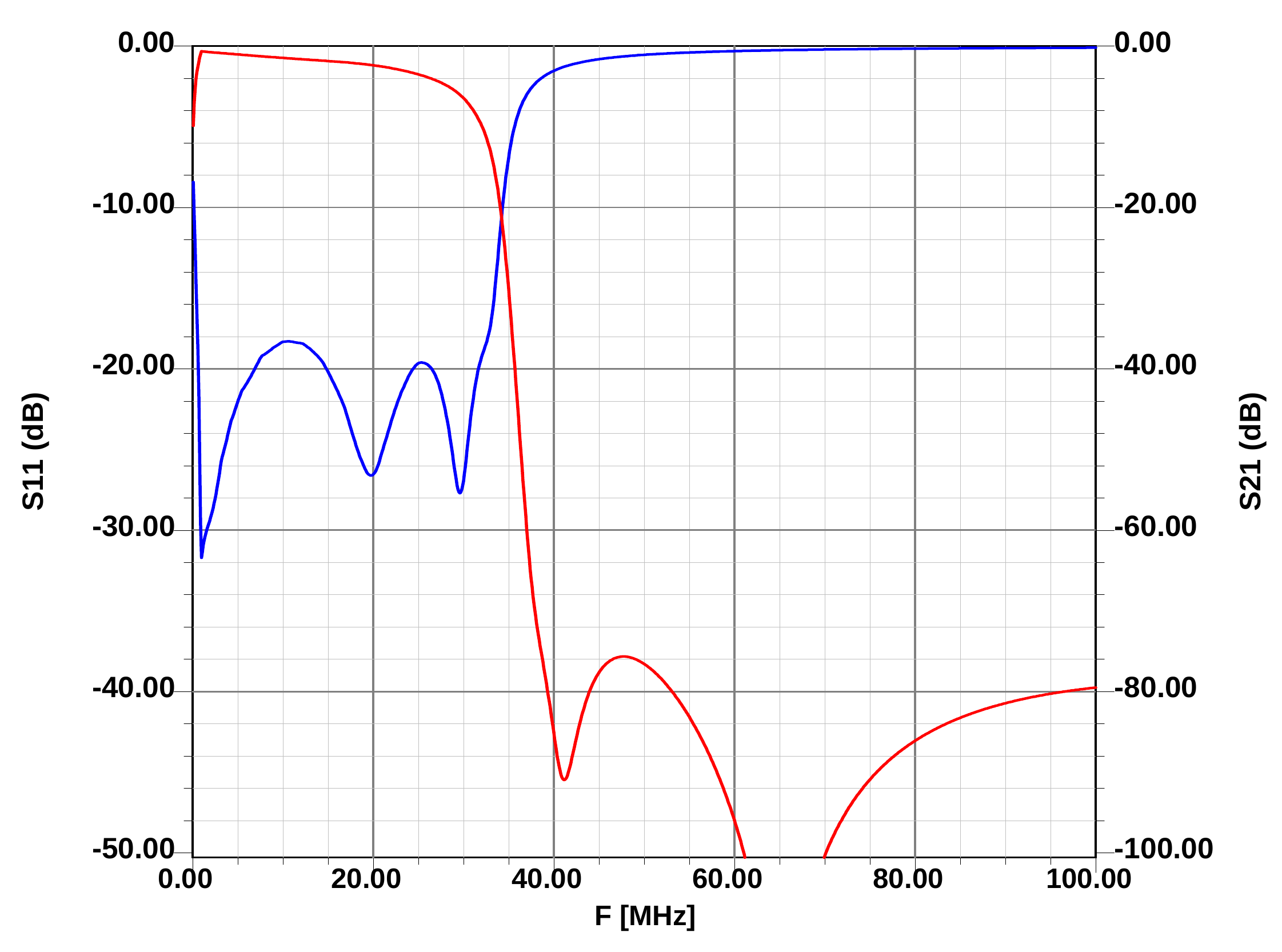}}
\caption{A simulation report of the lowpass filter design, including a unit gain buffer.} \label{fig:LPF_Sim}
\end{figure}

To conclude this section, we attach a photo of the analog board in Fig.~\ref{fig:mboard}. The splitting stage and the four channels are noticed in the layout. We point out the shielding holes between the RF paths, which we added to improve isolation between the mixer components. Table~\ref{table:devices} lists the devices we used with their manufacturer reference names. The
table allows to reproduce our design in a standard electronic
laboratory.

\ifTwoColumns \renewcommand{\FigWidth}{0.8\linewidth} \else
\renewcommand{\FigWidth}{0.5\linewidth} \fi
\begin{figure}
\centering \mbox {\includegraphics[width=\FigWidth]{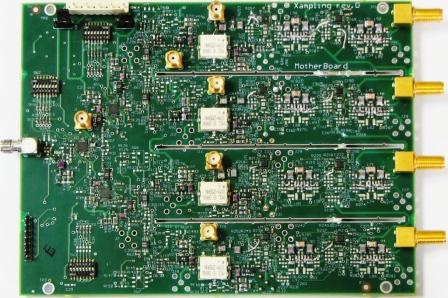}}
\caption{The analog board realizing four sampling channels. It consists of three stages:
splitting the analog input, mixing with four input periodic
waveforms (connectors on the rear), and lowpass filtering.} \label{fig:mboard}
\end{figure}

\renewcommand{\TableSize}{\scriptsize}
\renewcommand{\ra}[1]{\renewcommand{\arraystretch}{#1}}
\begin{table}\centering\caption{Device list}
\ra{1.2}{\TableSize
\begin{tabular}{@{}llll@{}}

\toprule

& Device & Reference & Manufacturer \\

\midrule

\multirow{8}{*}{\begin{sideways}\textbf{Analog
board}\end{sideways}}&
LNA & SPF5043 & RFMD\\
&Amplifier & Gali-21+ & Mini-Circuits \\
&Amplifier & Gali-84 & Mini-Circuits \\
&Attenuators & DAT-15R5-SN & Mini-Circuits \\
&RF-switch & HMC284MS8G & Hittite Microwave Corp.\\
&Mixer & SYM25-DHW & Mini-Circuits \\
&Filter amplifier & ADA4817 (single-ended) & Analog Devices\\
&Filter amplifier & ADA4932-1 (balanced) & Analog Devices\\

\midrule

\multirow{7}{*}{\begin{sideways}\textbf{Digital
board}\end{sideways}}&
Shift-register & MC10EP142MNG & ON Semiconductors \\
&VCO & ROS-2082-119+ & Mini-Circuits \\
&Synthesizer & ADF4106 & Analog Devices \\
&Crystal & 5597ASX3SVT & European Crystal Org. \\
&ECL clock splitter & ADCLK925 & Analog Devices \\
&Amplifier & Gali-21+ & Mini-Circuits \\
&CPLD & EPM570T144 & Altera \\

\bottomrule
\end{tabular}}
\label{table:devices}
\end{table}

\section{Digital Board}\label{sec:dboard}

\subsection{Description}

The digital board of our system is described schematically in Fig.~\ref{fig:digitalboard}(a). The main block is a SR of 96 bits at emitter-coupled-logic (ECL) technology, concatenating 12 packages of an 8-bit SR each. The initial value of the SR is
\begin{equation}\label{eq:initval}
  \textrm{Pattern} = \textsf{43\;A7\;A5\;D7\;96\;AB\;62\;B7\;2A\;B3\;5C\;AC},
\end{equation}
encoded in hexadecimal bytes. In order to load the register, we permit either parallel load of a fixed pattern, which is set manually by mechanical switches, or by programming the pattern through an on-board small field-programmable-gate-array (FPGA) device. The clock frequency of this device is set to 6 MHz to minimize the power consumption.

\ifTwoColumns \renewcommand{\FigWidth}{0.5\linewidth} \else
\renewcommand{\FigWidth}{0.5\linewidth} \fi
\begin{figure}
\centering \mbox {
\subfigure[]{\includegraphics[width=\FigWidth]{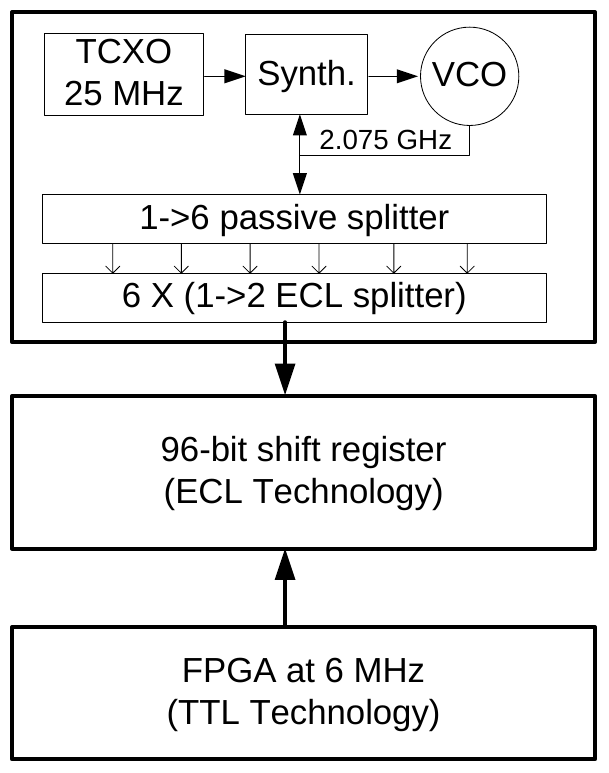}}
\subfigure[]{\includegraphics[width=\FigWidth]{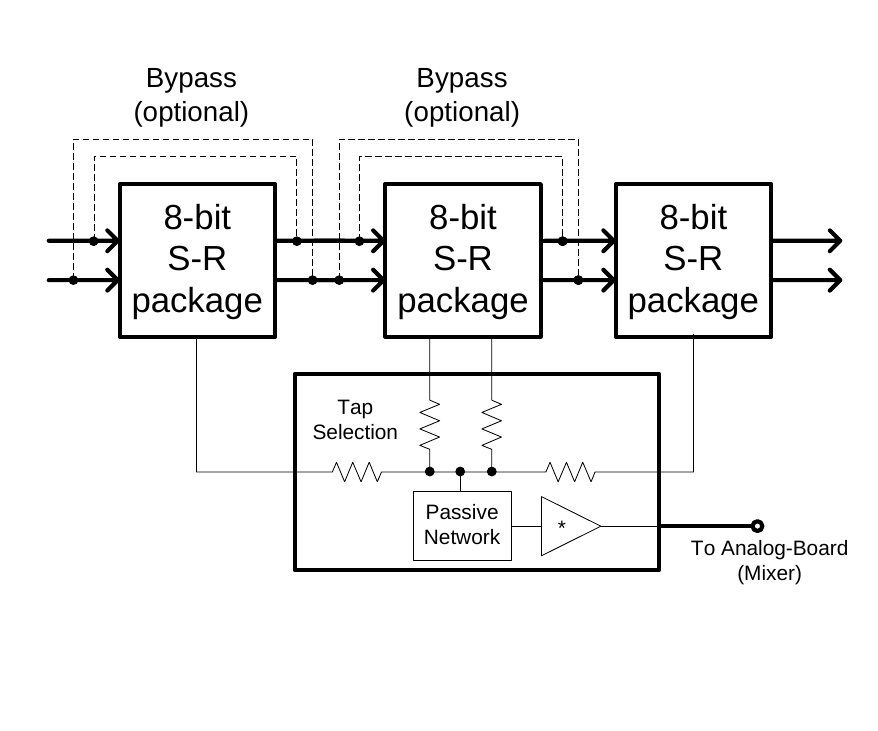}}}
\caption{Block diagram of the digital board is shown in (a). The tap selection and the bypass options are presented in (b).} \label{fig:digitalboard}
\end{figure}

Each analog channel receives a different tap of the SR, thus the waveforms $p_i(t)$ are shifted version of each other. To allow flexibility, the tap locations are not fixed, but rather have several configurations as set by the resistors in Fig.~\ref{fig:digitalboard}-b. The first 24 bits of the SR have four taps selection for $p_1(t)$. Similarly, the second waveform is derived from the next 24 bits and so on. To allow further flexibility, we allow to control the SR length by adding bypass options that can exclude the first or the second package in every triplet. In practice, we did not use the bypass options and connected each $p_i(t)$ to the 5th tap of the relevant series. The register length, the initial value (\ref{eq:initval}) and the taps locations are based on the theoretical study in \cite{ME09T2P,ME09EXRIP} and on the design considerations explained in the next section.

The clock network for the SR packages is derived from a 2.075 GHz sine waveform that is synthesized by locking a standard VCO to a 25 MHz temperature-compensated crystal oscillator (TCXO). The sine waveform is split to 6 identical sine waveforms using passive splitters. Then, we use lumped baluns to translate the single-ended sine waveform to the balanced ECL input followed by couple barrier diodes for course shaping to ECL levels. Fig.~\ref{fig:lumpeddiode} depicts the schematics of this stage. Finally, 6 ECL splitters generate 12 clock signals at valid levels. No further splitting is required, since each package of 8-bit SR is triggered by a single clock signal. The stability of the
clock network is the key to ensuring periodicity of the mixing
waveforms $p_i(t)$. We routed the clock signals in short
straight lines so as to avoid unintended time skews.

\ifTwoColumns \renewcommand{\FigWidth}{0.5\linewidth} \else
\renewcommand{\FigWidth}{0.5\linewidth} \fi
\begin{figure}
\centering \mbox {
\subfigure[]{\includegraphics[width=\FigWidth]{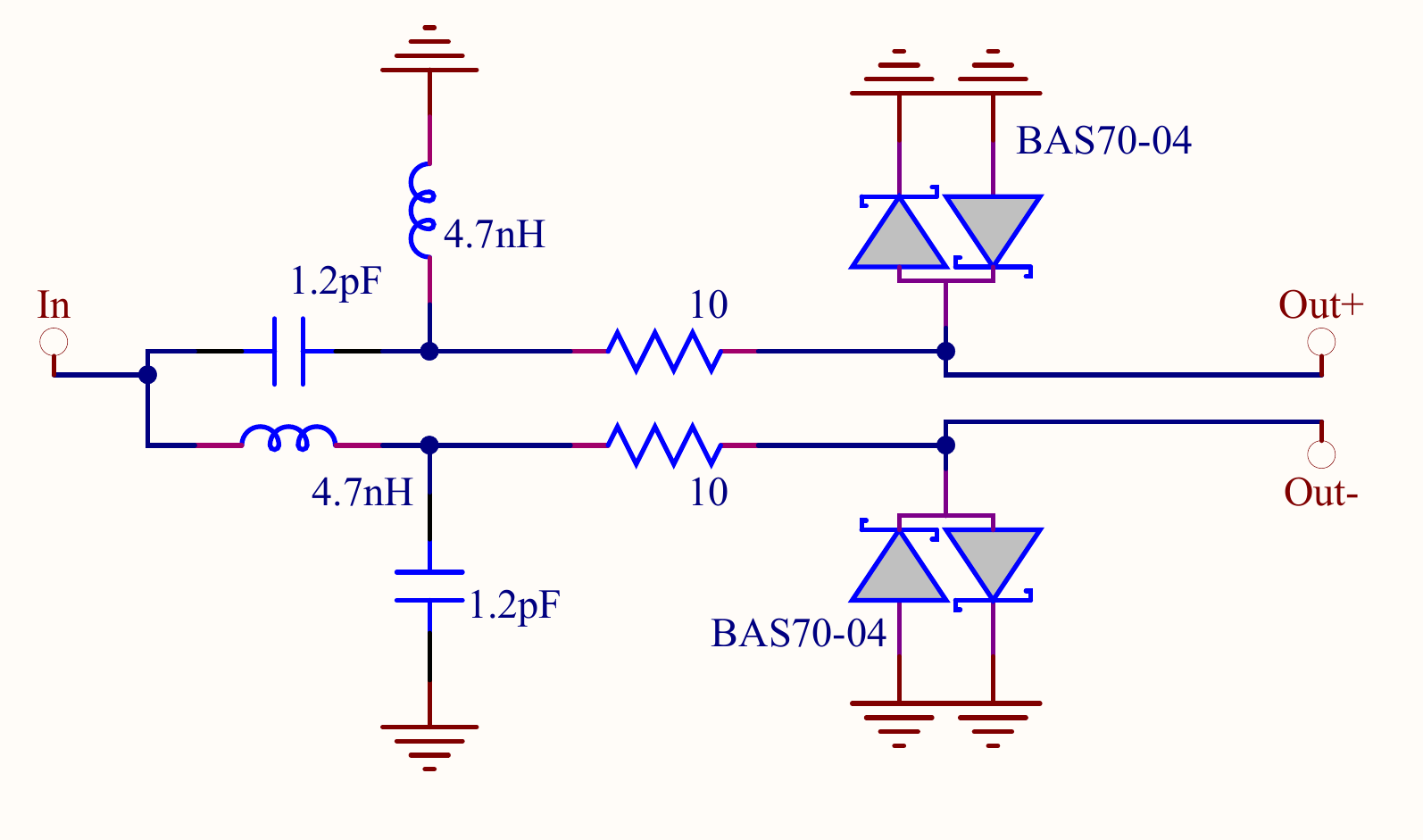}}
\subfigure[]{\includegraphics[width=\FigWidth]{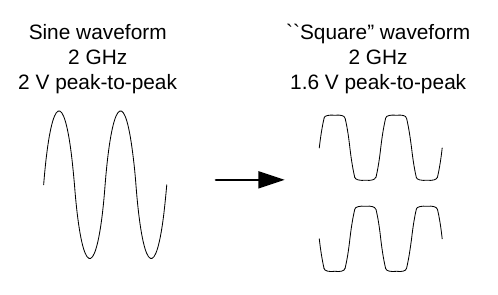}}}
\caption{The lumped balun (a) converts the single-ended sine waveform to a balanced waveform pair. The voltage barrier diodes collapse the sine to a square-like waveform (b), which matches the ECL standard. The output is further shaped by an ECL splitting device.} \label{fig:lumpeddiode}
\end{figure}

The FPGA interface to the SR consists of a bus of 96 signals for the parallel programming, a reset signal and a mode signal. On power-up, the mode signal switches the SR from programming to operational mode in which serial shifting occurs at 2.075 GHz rate. Since these control signals appear asynchronously, we double-buffer them to avoid metastability affects. Such buffering is not required for the bus, since the pattern is fixed while programming.

\subsection{Design considerations}

Realizing the SR turned out to be a circuit challenge. Conventionally, logic design is performed and simulated in software. The hardware implementation is then accomplished by programming the design in an FPGA device, which is a general purpose platform that can realize various logic structures and wire them internally. Working with FPGA has many advantages as it allows focusing on the logic structure, while relying on optimizations that are done by the manufacturer regarding amplitude levels, low skew on the clock network etc. We used this methodology for programming the SR on start-up. Unfortunately, this approach is not suited for realizing a high-rate SR. Popular FPGA devices do not stand a clock rate of $2.075$ GHz and premium devices are power consuming and very expensive. Therefore, we had to depart from conventional logic design and realize the SR using discrete devices.

The difficulty in implementing a discrete SR at 2.075 GHz is in satisfying the timing constraints. For correct functionality, the data must propagate from one package to the following such that it arrives sufficiently early before the next clock edge, \aka a setup time requirement. In addition, the data should not arrive too fast as otherwise it ruins the contents of the previous cycle, \aka a hold time constraint. The clock period of 480 picosecs renders these requirements very difficult to maintain, since delays due to the routing between the discrete devices are on the same order of the clock period. The delays affect both the data propagation and also imply uncertainty in the clock skew between the packages. Furthermore, the rising and falling durations of the logic signals are also on the same order. Any solution which is based on fine tuning of such short time delays is prone to errors; the wide literature on time mismatches in interleaved-ADCs is evidence to this difficulty, cf. \cite{huang2007blind,camarero2008mixed,ECM09,NM09,tsai2009correction,wang2006background}.

To overcome this challenge, our design is based on the MC10EP142MNG device, which implements an internally-wired 8 bits SR in
ECL technology. We exploited a specific property of that device. The datasheet reports on a maximal shifting frequency of 2.8 GHz (equivalent to a minimal clock period of 357 picosecs), and a delay from clock to output of about 670 picosecs. The meaning is that the manufacturer guarantees correct functionality in the internal register, up to a shifting rate of 2.8 GHz, though once the signal leaves the device the delay is almost doubled. Capitalizing on this difference is the key for satisfying the setup and hold requirements with no additional hardware. As illustrated in Fig.~\ref{fig:clocks}, since the delay of the last bit in each package is larger than the clock period, it reaches the next package one clock later than what would be normally expected in a standard FPGA design. This behavior seems erroneous and is likely to be reported as a violation of timing constraints by software tools. Fortunately, it follows from Fig.~\ref{fig:clocks} that we can view the large delay as an additional flop in between adjacent packages, with an equivalent clock-to-output delay of 670-480 = 190 picosecs. In other words, the data arrives about half a clock period before the next edge, in which case both the setup and time requirements are satisfied. To emphasize, the proposed solution requires no time-delay elements or synchronization mechanisms whatsoever. This allows a sign pattern of length $M=108$ bits to be realized by only 96 physical flops.

\ifTwoColumns \renewcommand{\FigWidth}{0.8\linewidth} \else
\renewcommand{\FigWidth}{0.5\linewidth} \fi
\begin{figure}
\centering \mbox {\includegraphics[width=\FigWidth]{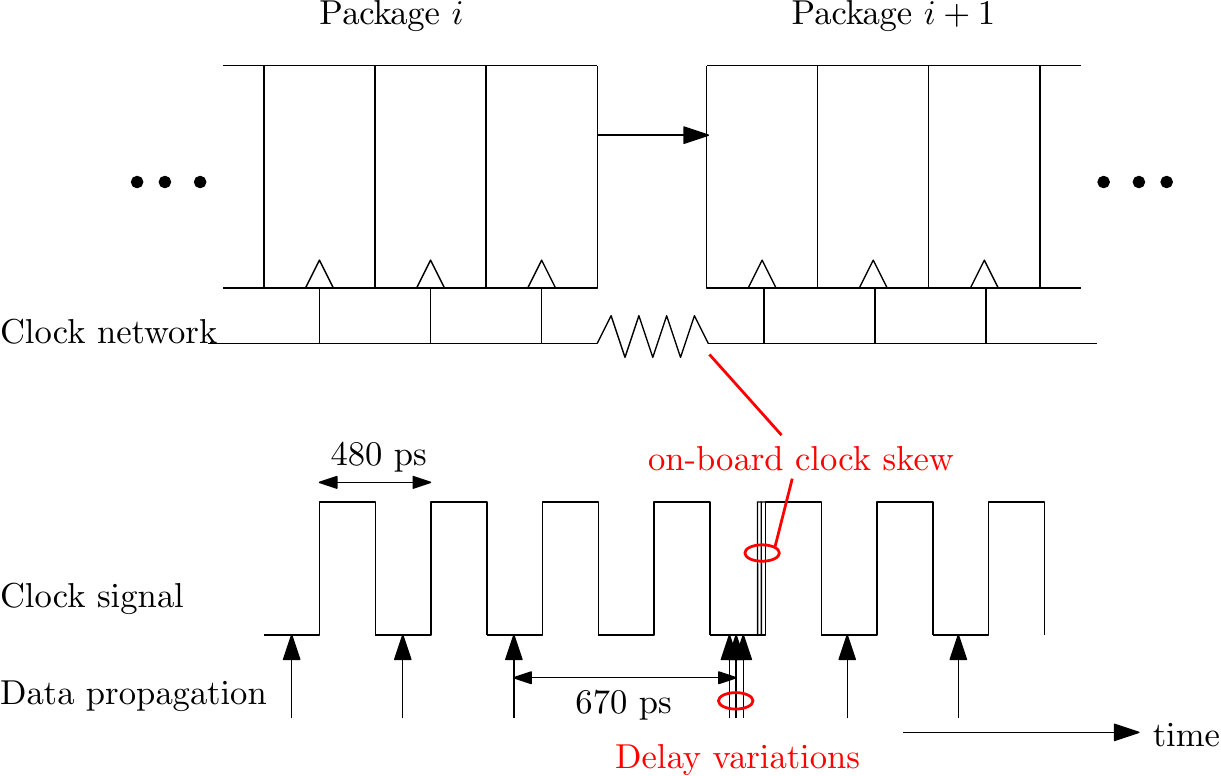}}
\caption{The interface of adjacent SR packages involves uncertainty on the clock network due to on-board routing. The timing diagram illustrates the propagation of data in the SR chain. Setup and hold constraints are satisfied internally by manufacturer, and externally by viewing the clock to output delay as an additional flop in between adjacent packages.} \label{fig:clocks}
\end{figure}

Our theoretical study in \cite{ME09EXRIP} quantifies the MWC performance for a given sign pattern, in terms of the expected restricted isometry property of a certain matrix. In practice, as we discuss in the experiment of Section~\ref{sec:exprsndr}, an important factor in choosing the sign pattern is balancing between the power of all the coefficients $c_{il}$. We therefore generated a set of tentative pattern options that stand the performance bounds of \cite{ME09EXRIP}, and then decided on (\ref{eq:initval}) according to the spectral appearance. The tap locations and the SR length are based on the same considerations.

Finally, we point out that the TCXO frequency balances two considerations: the basis frequency is low enough to mitigate possible interferences with the analog board, and on the other hand, it is high enough to reduce phase noise in the VCO. The TCXO datasheets specify 5 ppm maximal frequency deviation over a wide temperature range, which translates to a maximal frequency deviation of 10 kHz (over temperature) in the clock rate. In practice, the experiments in Section~\ref{sec:expr} quantify the clock jitter at standard room temperature and prove that the deviation is negligible.

A photo of the digital board appears in Fig.~\ref{fig:dboard}. The register chain is noticed, with the clock splitting network residing in the middle of the chain. The switches for controlling the sign pattern are located on the right side.

\ifTwoColumns \renewcommand{\FigWidth}{0.8\linewidth} \else
\renewcommand{\FigWidth}{0.5\linewidth} \fi
\begin{figure}
\centering \mbox {\includegraphics[width=\FigWidth]{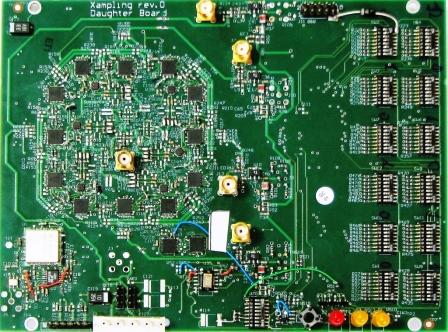}}
\caption{The digital board provides four sign-alternating periodic waveforms, which are derived from a single SR chain.} \label{fig:dboard}
\end{figure}

\section{Experiments}\label{sec:expr}

In this section, we describe several experiments that we conducted in order to verify the design. The last experiment demonstrates sub-Nyquist sampling in practice. The lab equipment that was used throughout includes: HP-8563E spectrum analyzer, Agilent Infiniium 54855A four-channel scope, HP-E4432B signal generator, Agilent 8753E network analyzer and HP 436A power meter.

\subsection{Periodicity and sign waveforms}\label{sec:exprsign}

The first test aims at verifying the periodicity of $p_i(t)$ which is the key for the MWC operation.

\begin{figure}[h]
\centering \mbox {
{\includegraphics[width=0.5\linewidth]{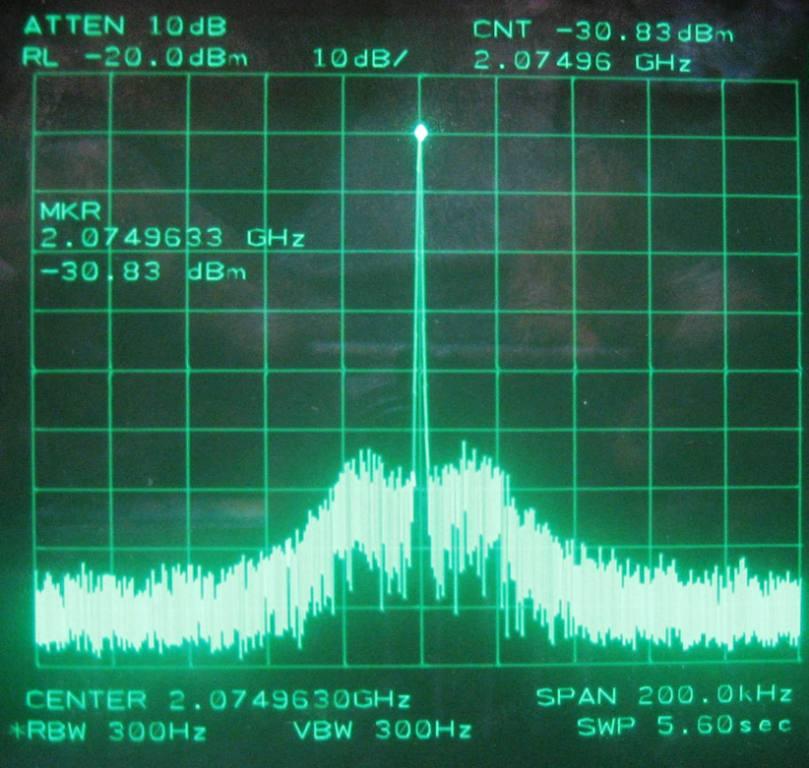}}}
\caption{The central frequency of the VCO is measured as $2.07496$ GHz. The noise hill-like shapes surrounding the center frequency are commonly referred to as the phase noise of the VCO.} \label{fig:exprvco}
\end{figure}

The periodicity of the mixing waveforms is dictated by the accuracy of the clock network, which is derived from the VCO. In Fig.~\ref{fig:exprvco}, we observe a 200 kHz span around the sine output of the VCO. The figure implies a phase noise of less than -52 dBc at 20 kHz aside the carrier, and less than -70 dBc at 60 kHz aside. A VCO with this phase noise is considered steadily locked for normal applications of frequency conversion.

\ifTwoColumns \renewcommand{\FigWidth}{0.75\linewidth} \else
\renewcommand{\FigWidth}{0.5\linewidth} \fi

\begin{figure}[h]
\centering \mbox {
\subfigure[]{\includegraphics[width=0.40\linewidth]{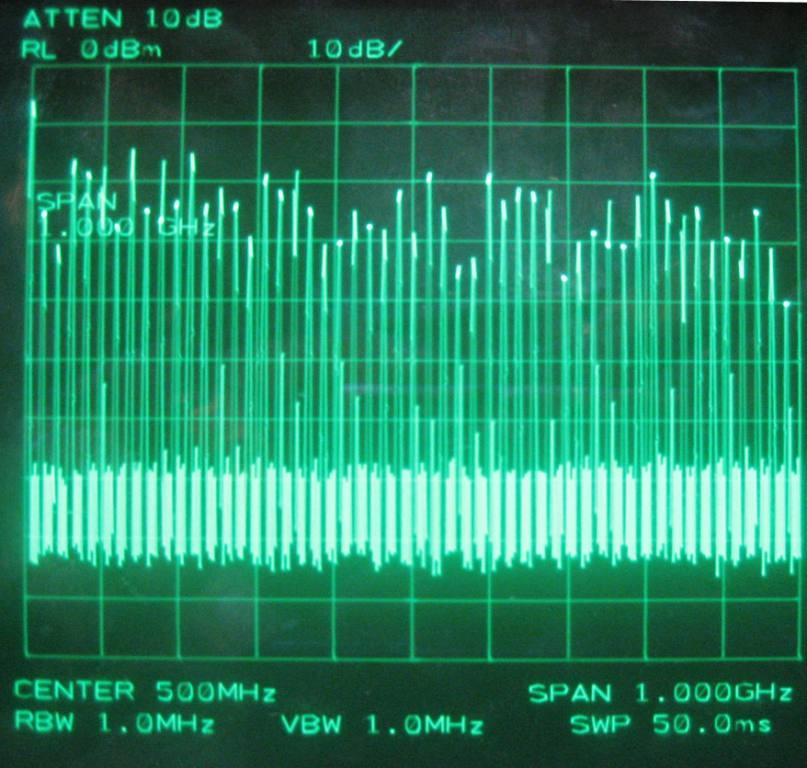}}
\subfigure[]{\includegraphics[width=0.40\linewidth]{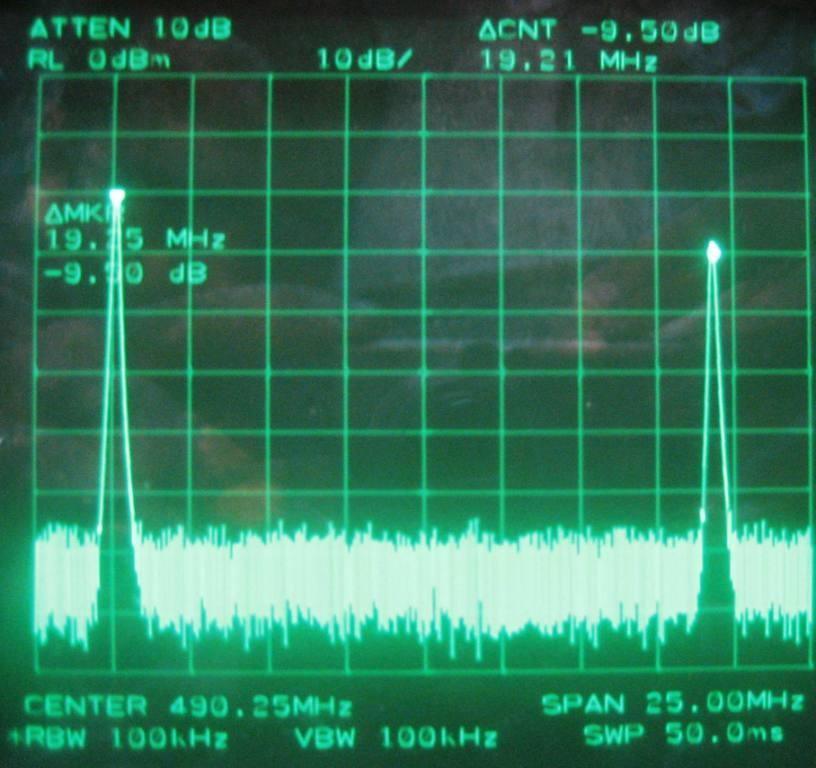}}}
\caption{The spectrum of a sign-alternating periodic waveform (a), and the spacing (b) between consecutive
Diracs.} \label{fig:exprwavefreq}
\end{figure}

To further verify the periodicity of the mixing waveforms, we observed the outputs of the digital board in a spectrum analyzer. Fig.~\ref{fig:exprwavefreq} depicts the spectrum of $p_i(t)$, which consists of
equalispaced Diracs, namely highly concentrated energy peaks, as expected for periodic waveforms. The spacing between the Diracs is measured as 19.212 MHz, which validates the design choice $f_p$ of Table~\ref{table:param}. The Dirac spectral lines appear steady, ensuring the periodicity of
$p_i(t)$. Note the balanced power levels of the Diracs comprising $p_i(t)$.

\ifTwoColumns \renewcommand{\FigWidth}{0.5\linewidth} \else
\renewcommand{\FigWidth}{0.5\linewidth} \fi

Finally, we observed the time-domain appearance of $p_i(t)$ in Fig.~\ref{fig:exprwavetime}. A zoom on a short time interval reveals that the  waveforms $p_i(t)$ are far from nice rectangular transitions on the Nyquist grid. However, since periodicity is the only essential requirement of the MWC, the nonideal time-domain appearance has no effect in practice.

\ifTwoColumns \renewcommand{\FigWidth}{0.45\linewidth} \else
\renewcommand{\FigWidth}{0.5\linewidth} \fi
\begin{figure}[h]
\centering \mbox {
\subfigure[]{\includegraphics[width=\FigWidth]{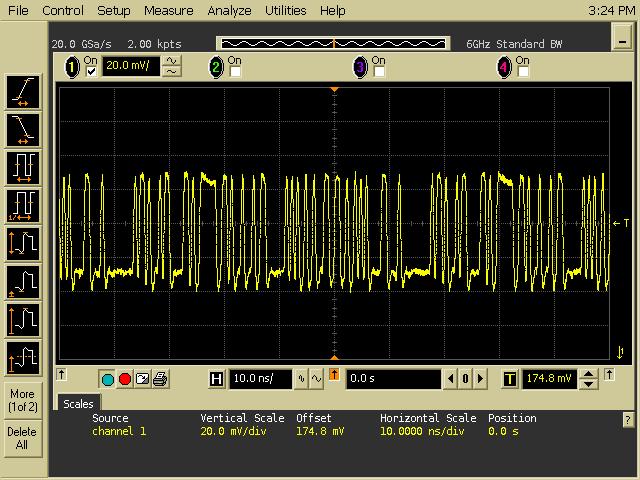}}
\subfigure[]{\includegraphics[width=\FigWidth]{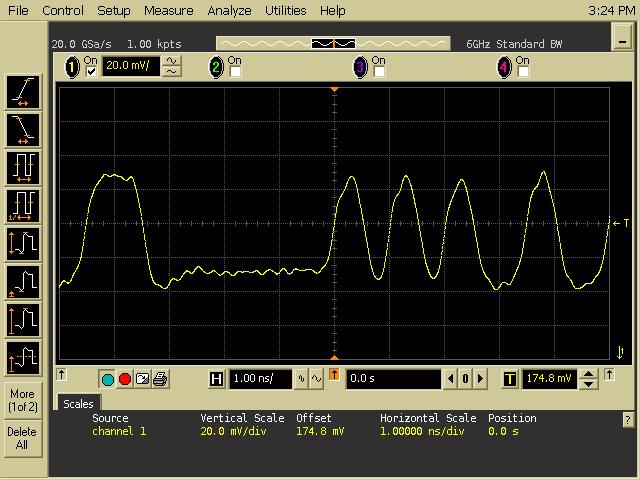}}}
\caption{A sign-alternating waveform at 2 GHz (a), and a zoom on a short time interval (b).} \label{fig:exprwavetime}
\end{figure}

\subsection{Power level, SNDR and dynamic range}\label{sec:exprsndr}

The purpose of this test is to verify that the outputs stand the required power level (\ref{eq:ampreq}) and SNDR (\ref{eq:snrreq}).

We synthesized a pure sinusoid signal at the system input, with 92.4 MHz frequency and -45~\dbm~input power. The on-board attenuators were set to 0 dB. Fig.~\ref{fig:expramp} displays the four outputs, as measured by the four-channel scope (with 50 ohm impedance input terminals). The amplitude levels are all around 1 Vptp, or equivalently 4 \dbm. A more accurate measurement, using a power meter, revealed that the output power is about 5 \dbm. It follows that the minimal input power satisfying (\ref{eq:ampreq}) is -55 \dbm, conforming with (\ref{eq:pinamp}).  As explained earlier, this test assures (\ref{eq:ampreq}) over all input powers higher than -55 \dbm, as long as setting the attenuators according to Fig.~\ref{fig:sinad}(b). Note, that the result also validates our modification of the mixer gain from -6 dB to -16 dB.

\ifTwoColumns \renewcommand{\FigWidth}{0.5\linewidth} \else
\renewcommand{\FigWidth}{0.5\linewidth} \fi
\begin{figure}[h]
\centering \mbox {
\includegraphics[width=\FigWidth]{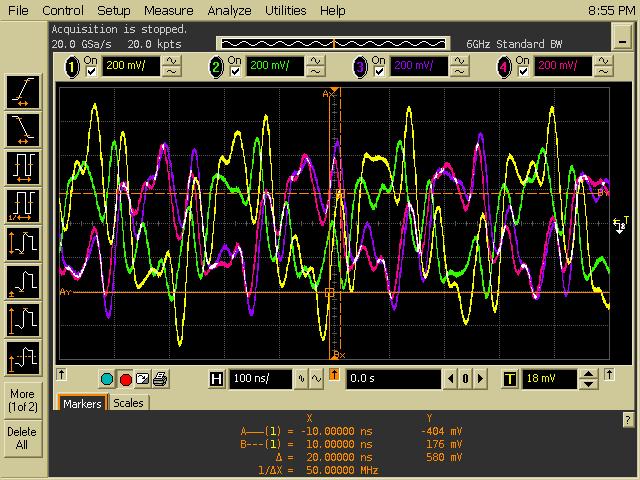}}
\caption{System outputs acquired by a four-channel scope.} \label{fig:expramp}
\end{figure}

\begin{figure}[h]
\centering \mbox {
\subfigure[]{\includegraphics[height=26mm]{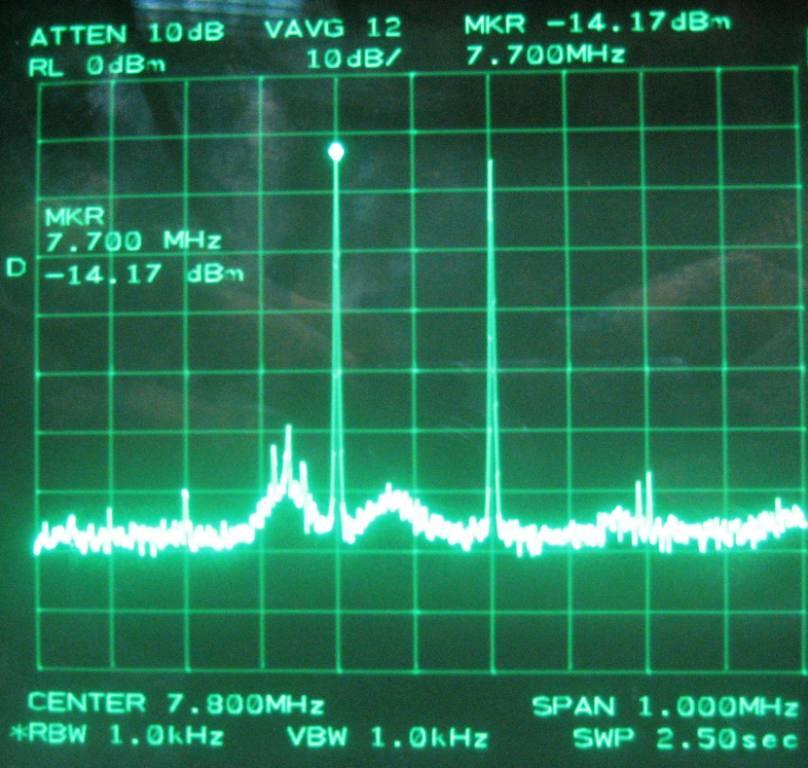}}
\subfigure[]{\includegraphics[height=26mm]{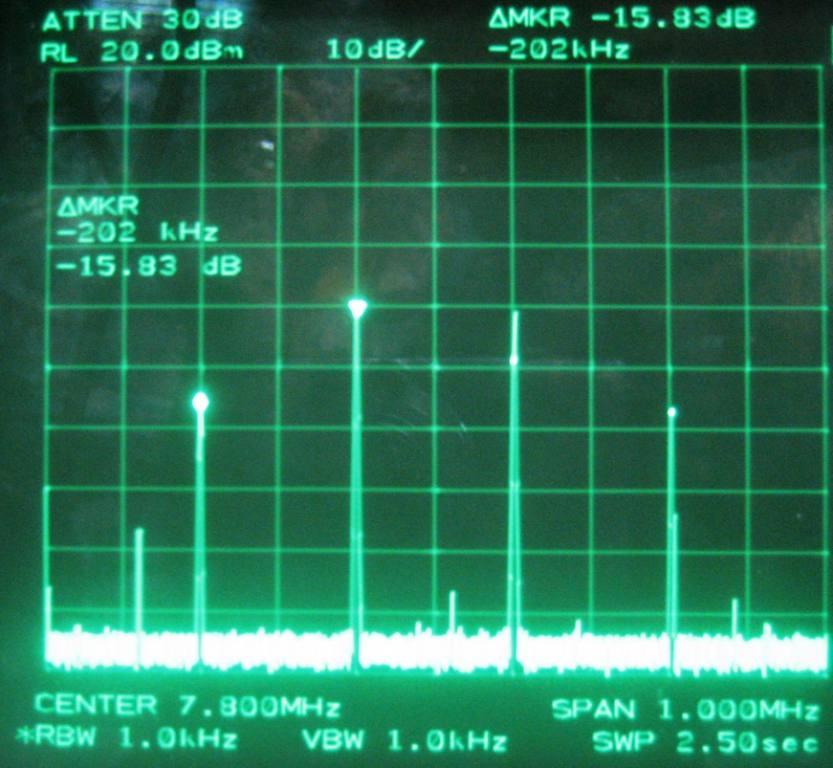}}
\subfigure[]{\includegraphics[height=26mm]{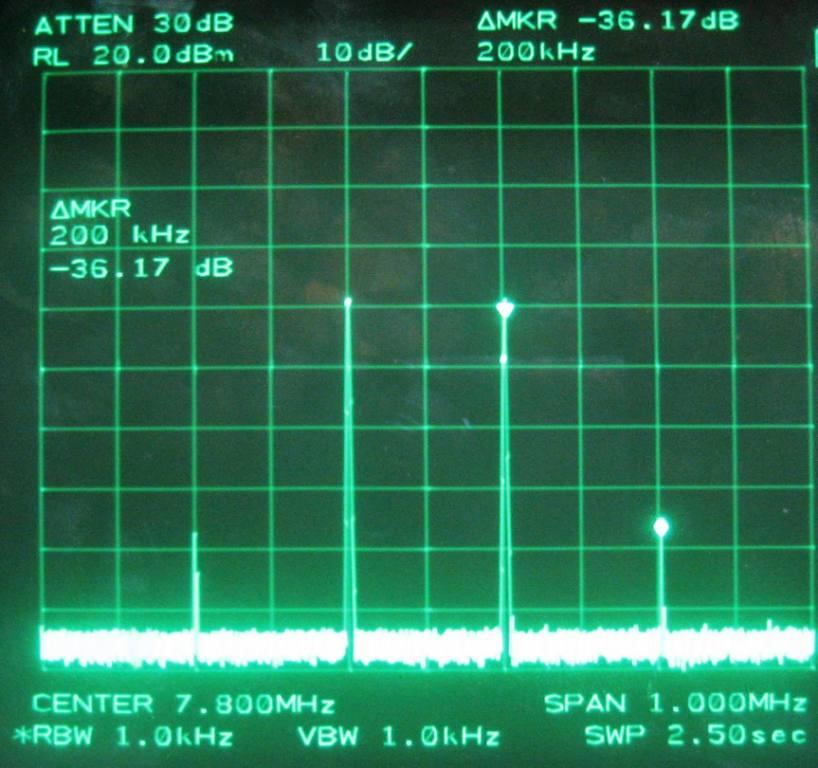}}}
\caption{Output spectrum for two equal-power input tones. Panel (a) measures the SNR for low input power. Panels (b) and (c) verify the nonlinear distortion levels for high signal energies. The noise floor in these panels is attributed to the spectrum analyzer. In (c), an external 20 dB attenuator is connected in series to the system output.} \label{fig:exprtwotones}
\end{figure}

To check the SNDR requirement (\ref{eq:snrreq}), we used an 1:2 RF combiner to sum two signal generators to the board input, with carriers 146 MHz and 145.8 MHz and equal powers. These inputs convolve with the 137 MHz Dirac of $p_i(t)$. Fig.~\ref{fig:exprtwotones}(a) displays the spectrum at the output of the first channel for $-55$ dBm input powers. The original tones, with the 200 kHz spacing, are noticed, with -14.2 dBm output power per tone. The noise level is -78 dBm at 1 kHz resolution bandwidth (RBW), which is amount to -33 dBm, when translating to the 33 MHz filter bandwidth. This verifies 18.8 dB SNR. Fig.~\ref{fig:exprtwotones}(b) shows the results for input powers of -5 dBm, where the attenuators are set to -15.5 dB. The 3rd order nonlinear distortions are 200 kHz aside the desired signals, with 15.83 dB signal to distortion ratio. These two measurements affirm a 50 dB dynamic range.

\ifTwoColumns \renewcommand{\FigWidth}{0.32\linewidth} \else
\renewcommand{\FigWidth}{0.5\linewidth} \fi

To verify the performance of the X-ADC architecture, we used the same experiment with -35 dBm input powers. The attenuators were set to according to Fig.~\ref{fig:sinad}(b), namely -10.5 dB and -15.5 dB, respectively. Fig.~\ref{fig:exprtwotones}(c) demonstrates 36.2 dB signal to distortion ratio, which implies ENOB=5.7 bits according to (\ref{eq:enob}).

\subsection{Filter response}

We measured the lowpass filter response from the IF terminal of the mixer to the single-ended output. Fig.~\ref{fig:exprfilter} affirms high attenuation of about 70 dB in the stopband. Note the smooth curve of both $S_{21}$ and $S_{11}$ parameters, which motivated splitting the filter design into two stages, as explained in the design considerations of Section~\ref{sec:mboard}. The nonflat frequency response in the passband, namely till 33 MHz, is compensated digitally \cite{Yilun09}.

\ifTwoColumns \renewcommand{\FigWidth}{0.4\linewidth} \else
\renewcommand{\FigWidth}{0.5\linewidth} \fi
\begin{figure}[h]
\centering \mbox {
\subfigure[]{\includegraphics[width=\FigWidth]{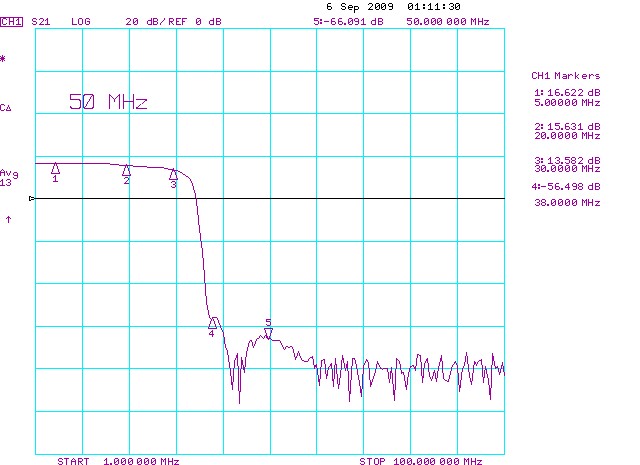}}
\subfigure[]{\includegraphics[width=\FigWidth]{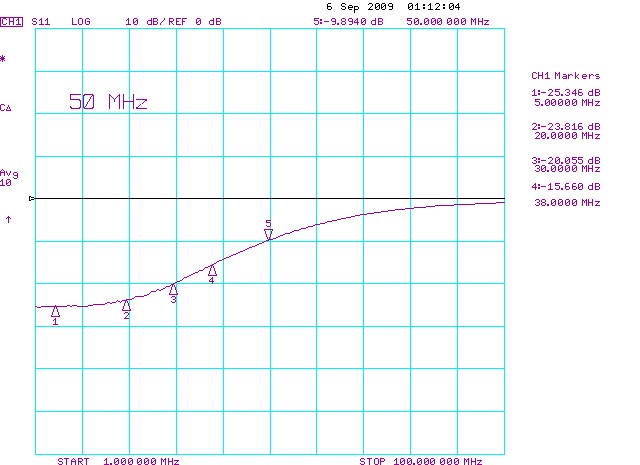}}}
\caption{Filter response: (a) forward $S_{21}$ and (b) input return loss $S_{11}$.} \label{fig:exprfilter}
\end{figure}

\subsection{Sub-Nyquist sampling}

\begin{figure*}
\centering \mbox {
\subfigure[]{\includegraphics[height=35mm]{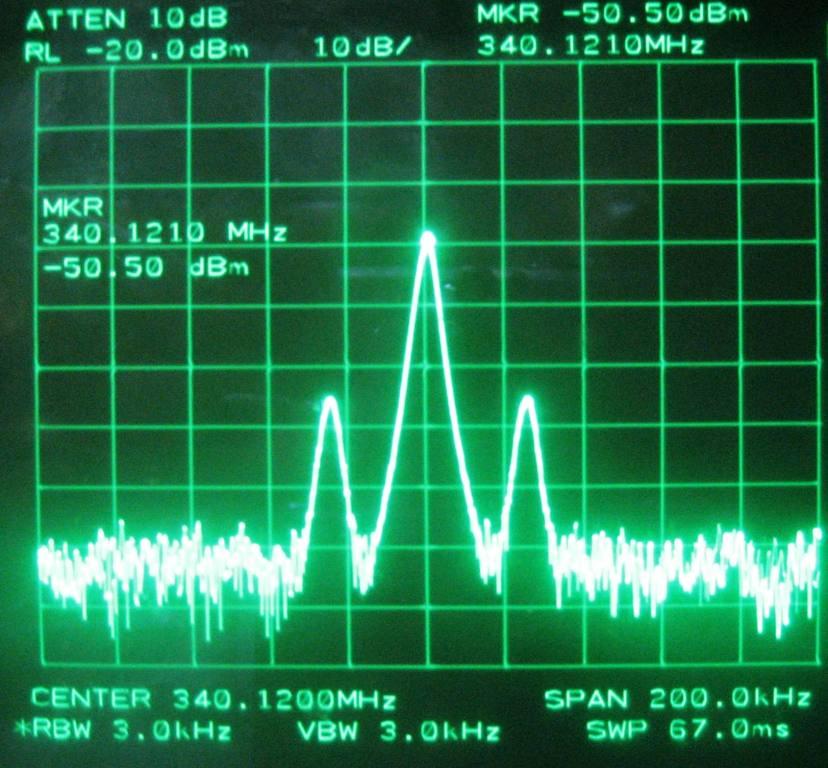}}
\subfigure[]{\includegraphics[height=35mm]{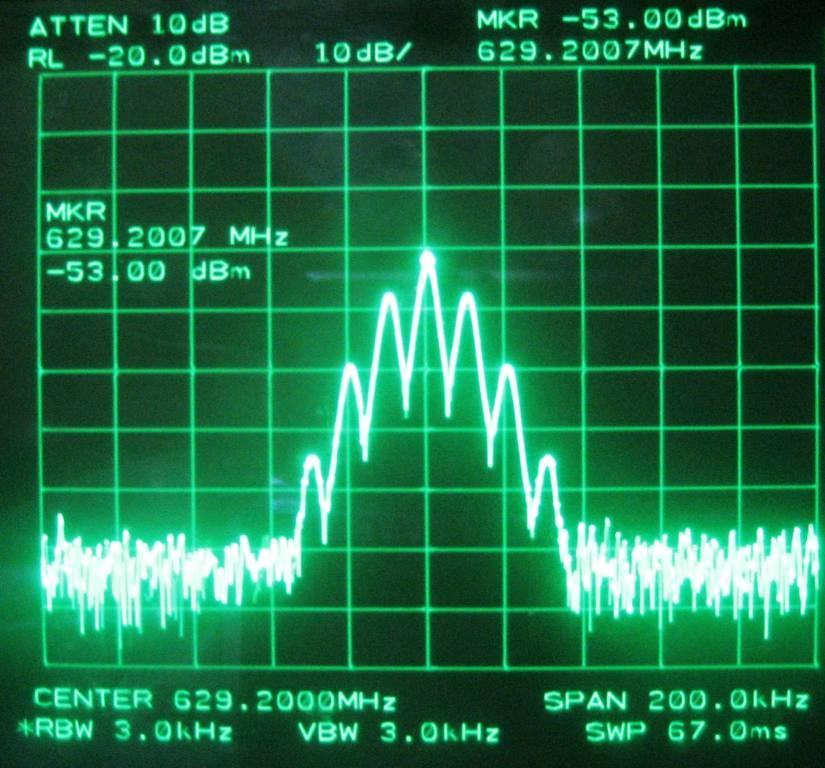}}
\subfigure[]{\includegraphics[height=35mm]{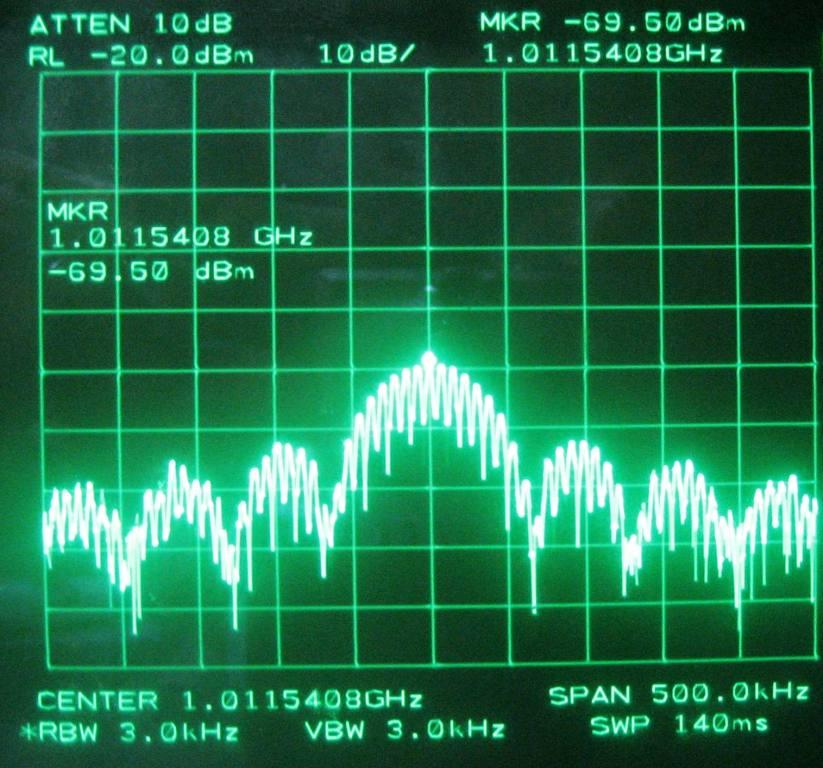}}}
\centering \mbox {
\subfigure[]{\includegraphics[height=35mm]{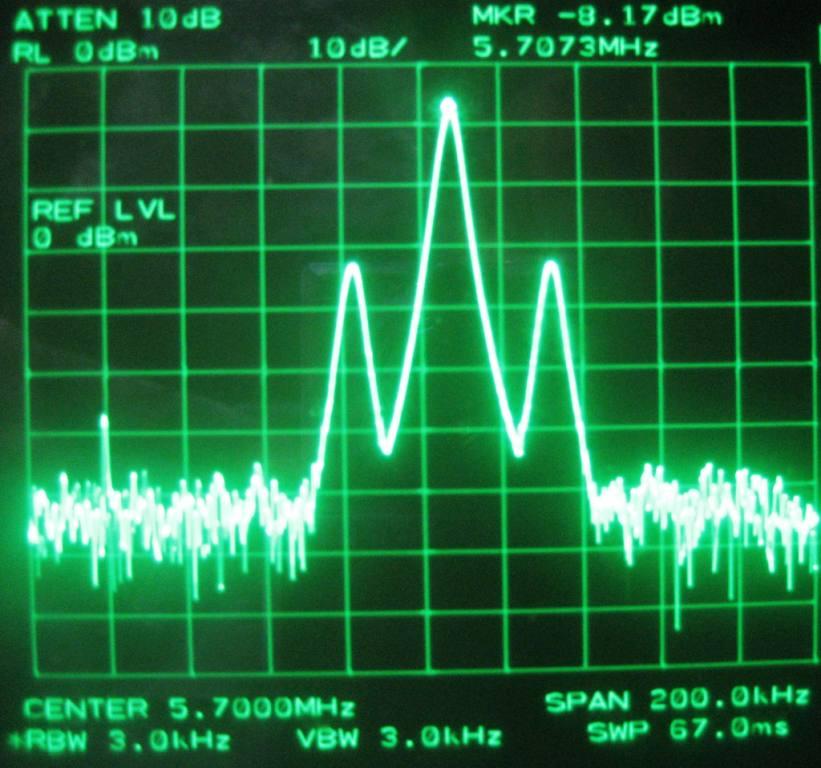}}
\subfigure[]{\includegraphics[height=35mm]{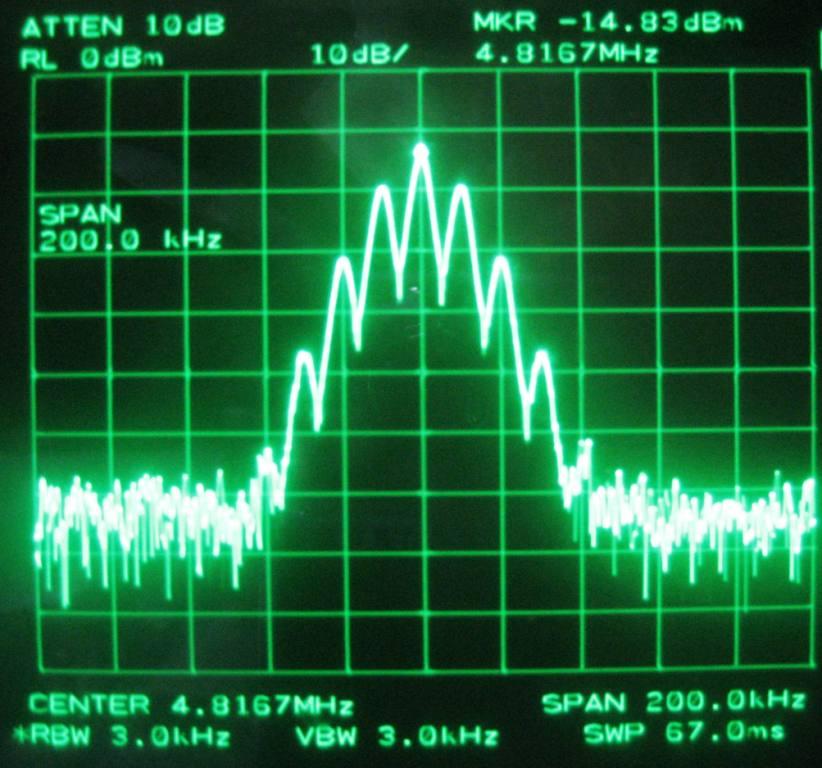}}
\subfigure[]{\includegraphics[height=35mm]{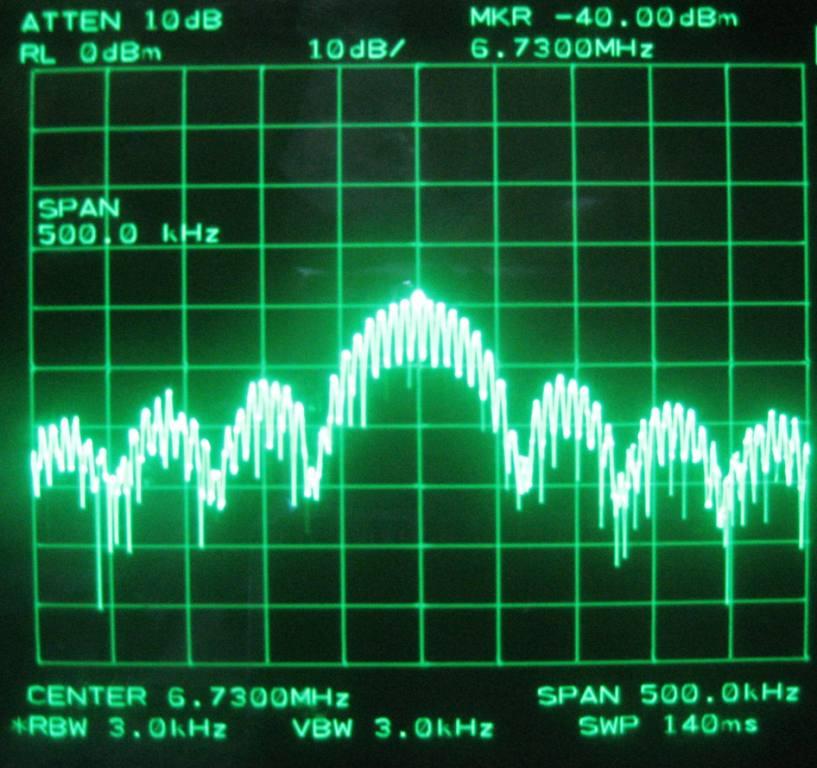}}}
\caption{Panels (a),(b),(c) plot the frequency shape of the AM, FM and PAM input signals, respectively. The sub-Nyquist outputs (first channel) are depicted beneath each input panel, affirming correct mixing operation, as supported by computations.}\label{fig:am}
\end{figure*}

In order to prove the sub-Nyquist sampling capability of our design, we generated three signal sources: amplitude-modulation (AM), frequency-modulation (FM) and pulse-amplitude modulation (PAM). The AM signal was positioned at $f_c=340.12$ MHz with 25 kHz sine amplitude and 10\% modulation depth. The FM signal was synthesized with $f_c=629.2$ MHz and 10 kHz sine frequency modulation depth. For the PAM signal, we used carrier $f_c=1011.54$ MHz, pulse period of 132 $\mu$secs and 16 $\mu$sec depth. The spectrum of the input is shown in Fig.~\ref{fig:am}, panels (a)-(c). The spectrum of the first sub-Nyquist output is plotted in (d)-(f), respectively. Evidently, the mixing shifts the entire wideband spectrum to low frequencies, while maintaining an accurate shape of the input.

To verify the frequency positions at the system output, we recall the mixture procedure as illustrated in Fig.~\ref{fig:mixtures}. For a carrier frequency $f_c$, let
\begin{equation}\label{fig:carr1}
  a = \textrm{rem}\left(\frac{f_c}{f_p}-\frac{1}{2}\right),
\end{equation}
where $\textrm{rem}(x)=x-\lfloor x \rfloor$ removes the integer part of $x$. The value of $a$ is the relative position of the carrier $f_c$ within the spectrum slice it resides. The MWC aliases all spectrum slices to the interval $[-f_p/2, f_p/2]$. Therefore, at the output the carrier $f_c$ is aliased to
\begin{equation}\label{fig:carr2}
  f_{\textrm{alias}} = \left|f_p\left(a-\frac{1}{2}\right)\right|.
\end{equation}
According to (\ref{fig:carr1})-(\ref{fig:carr2}) the AM, FM and PAM signals in our experiment are expected to alias to 5.71 MHz, 4.82 MHz and 6.73 MHz, respectively. These numbers are verified in Fig.~\ref{fig:am}.  
We note that two similar aliasing shapes appear around $f_p\pm f_{\textrm{alias}}$, since $f_s=3f_p$ in our design. The fact that the signal shape is unaltered is the key for enabling the subsequent digital processing, which were developed in \cite{ME09T2P,XamplingPractice}. The integration between the RF front-end and the digital stage is the subject of a forthcoming publication.

\section{Conclusions}\label{sec:conc}

We presented a circuit realization of the modulated wideband converter -- a sub-Nyquist sampler. Our design can handle wideband inputs and sample them at a fraction of the Nyquist rate. The theoretical study underlying the present work assumes a multiband model, in which the wideband input consists of several transmissions with unknown carrier positions. The circuit work follows the theoretical guidelines and adds another layer of considerations that were needed in order to realize the converter approach.

We have chosen to implement a specific configuration of the converter, in which the Nyquist rate of the input is around 2 GHz and the spectrum occupancy reaches 120 MHz. The sampling rate is as low as 280 MHz and slightly above the lowest possible rate for multiband signals with unknown carrier positions.

In addition to presenting the sub-Nyquist sampler, we put an emphasis on two circuit challenges: mixing a signal with multiple sinusoids and generating periodic waveforms with transients at the Nyquist rate. In order to achieve the desired effects, we employed standard devices, modified their specifications due to the nonordinary usage and added auxiliary circuitry accordingly. Further investigation of these circuit structures, beyond the current application of sub-Nyquist sampling, may assist in developing alternative solutions for these tasks. Future work will  report on a chip-level implementation of the converter.

\section*{Acknowledgment}
The authors would like to thank the technical teams of both the Communication Laboratory (ComLab) and the High-Speed Digital Systems Laboratory (HSDSL) at the technion for investing time and resources in this project. We also thank Moshe Namer, Dr. Avraham Saad and Prof. Moshe Nazarathy for the fruitful discussions on the preliminary architecture, and Shraga Krauss and Dr. Yossi Hipsh for many insightful comments.

\bibliographystyle{IEEEtran}
\bibliography{IEEEabrv,moshikobib}

\end{document}